\documentclass{article}
\usepackage{microtype}
\usepackage{graphicx}
\usepackage{subcaption}
\usepackage{booktabs} 

\usepackage[hyphens]{url}
\usepackage{hyperref}
\usepackage{enumitem}
\usepackage{placeins}

\usepackage[accepted]{icml2026}
\usepackage{amsmath}
\usepackage{amssymb}
\usepackage{mathtools}
\usepackage{amsthm}
\usepackage{caption}

\theoremstyle{plain}

\theoremstyle{definition}

\theoremstyle{remark}

\usepackage{makecell} 
\usepackage{graphicx}
\usepackage{textcomp}
\usepackage{booktabs}
\usepackage{longtable}
\usepackage[dvipsnames]{xcolor}
\def\BibTeX{{\rm B\kern-.05em{\sc i\kern-.025em b}\kern-.08em
    T\kern-.1667em\lower.7ex\hbox{E}\kern-.125emX}}
\usepackage{hyperref}
\usepackage{algorithm,algorithmic}
\usepackage{multirow}
\usepackage{xspace}
\usepackage{comment}
\usepackage{pgfplots}
\usepackage{framed, xcolor}
\usepackage{subcaption}
\usepackage{tabularx}
\usepackage{float}

\usepackage{pgfplotstable}
\usepackage{tikz}
\usepgfplotslibrary{statistics}
\usepackage{stmaryrd}
\pgfplotsset{compat=1.18}
\usetikzlibrary{pgfplots.fillbetween}
\pgfplotsset{compat=1.18}

\newcommand{\sSy}{\textit{m}\xspace}
\newcommand{\sW}{n\xspace}
\newcommand{\sCu}{\textit{j\xspace}}
\newcommand{\spacing}{\textit{step}\xspace}
\newcommand{\cue}{\textit{cue}\xspace}
\newcommand{\sign}{\textit{reply}\xspace}
\newcommand{\lo}{\textit{K}\xspace}
\newcommand{\col}{\mathcal{D}\xspace}
\newcommand{\doc}{\textit{D}\xspace}
\newcommand{\chunk}{\textit{$D_i$}\xspace}
\newcommand{\word}{\textit{d}\xspace}

\newcommand{\ws}{\textit{W}\xspace}
\newcommand{\w}{\textit{w}\xspace}
\newcommand{\syllable}{\textit{s}\xspace}
\newcommand{\cinr}{\textit{t}} 

\icmltitlerunning{Data Provenance Auditing of Fine-Tuned Large Language Models with a
Text-Preserving Technique}

\begin{document}
\twocolumn[
  \icmltitle{Data Provenance Auditing of Fine-Tuned Large Language Models with a Text-Preserving Technique}




  \begin{icmlauthorlist}
    \icmlauthor{Yanming Li}{inria,insa,ups}
    \icmlauthor{Cédric Eichler}{insa,uo,inria}
    \icmlauthor{Nicolas Anciaux}{inria,insa,ups}
    \icmlauthor{Alexandra Bensamoun}{ups}
    \icmlauthor{Lorena Gonzalez-Manzano}{cosec}
    \icmlauthor{Seifeddine Ghozzi}{ensta}
    
  \end{icmlauthorlist}

  \icmlaffiliation{insa}{INSA CVL, Bourges, France}
  \icmlaffiliation{uo}{Université d'Orléans, Orléans, France}
  \icmlaffiliation{inria}{Petscraft, Inria, Palaiseau, France}
  \icmlaffiliation{ensta}{ENSTA, Institut Polytechnique de Paris, Palaiseau, France}
  \icmlaffiliation{ups}{Université Paris-Saclay,  Palaiseau, France}
  \icmlaffiliation{cosec}{Universidad Carlos III de Madrid, Madrid, Spain}

  \icmlcorrespondingauthor{Cédric Eichler}{cedric.eichler@inria.fr}

  \icmlkeywords{Watermarking, LLM, invisible unicode}

  \vskip 0.3in
]



\printAffiliationsAndNotice{}  

\begin{abstract}
We propose a system for marking sensitive or copyrighted texts to detect their use in fine-tuning large language models under black-box access with statistical guarantees.
Our method builds digital “marks” using invisible Unicode characters organized into (“cue”, “reply”) pairs.
During an audit, prompts containing only “cue” fragments are issued to trigger regurgitation of the corresponding “reply”, indicating document usage.
To control false positives, we compare against held-out counterfactual marks and apply a ranking test, yielding a verifiable bound on the false positive rate.
Empirically, we obtain a true positive rate of 96.7\% at 0\% false positive rate and reply regurgitation rates exceeding 28\% per document with only 40 (4\%) watermarked documents.
The approach is minimally invasive, scalable across many sources, robust to standard processing pipelines, and achieves high detection power even when marked data is a small fraction of the fine-tuning corpus.

\end{abstract}

\section{Introduction} \label{sec:intro}

Large language models (LLMs) are trained and fine-tuned using vast collections of text, but the fine-tuning stage, based on task-specific often proprietary corpora, raises particularly salient questions about provenance, as the composition of these datasets is typically not disclosed. 
Fine-tuning data frequently consists of high-value content, including personal or copyrighted works (e.g., resumes for recruitment models, literary texts for creative assistance), whose limited size and targeted nature can disproportionately shape a model's downstream behavior.
This opacity leaves data owners and authors without effective means to determine whether their content has been used to fine-tune a deployed model or chatbot, creating challenges for privacy, intellectual property, and data governance that legal and policy instruments alone struggle to address. Beyond adversarial settings, provenance guarantees can also be meaningful in trusted deployment scenarios, such as fine-tuning-as-a-service, where a cooperative provider enables post-hoc auditing to verify whether marked documents were included in training.

We consider the following research question: given only black-box access to a fine-tuned LLM, can we reliably determine whether a specific corpus of documents belonging to a given rights holder was included in the fine-tuning set? We assume a minimal access setting with only prompt–response interaction, consistent with typical domain-specific LLM deployments, such as customer-service chatbots obtained through fine-tuning. 

Existing technical approaches fall short of providing reliable text-preserving provenance evidence. Training data can sometimes reappear verbatim in model outputs~\cite{carlini2022quantifying}, but such occurrences are hard to elicit and do not provide statistically controlled evidence\footnote{See an external analysis by Nicholas Carlini (2025), which explicitly advises against using these results for copyright litigation (\href{https://nicholas.carlini.com/writing/2025/privacy-copyright-and-generative-models.html}{link}).}. Membership inference attacks (MIAs) have been increasingly adapted to data provenance in LLMs, but their detection power~\cite{duan2024membership,DBLP:conf/satml/MeeusSJFRM25} and evidentiary reliability~\cite{10992516} remain limited and contested in post-hoc settings. Prior copyright-protection techniques for text~\cite{du2025sok} are typically invasive, relying on visible or semantically unnatural text modifications that are incompatible with sensitive or high-quality content.

We introduce a minimally invasive method for auditing fine-tuning data usage that enables statistically sound membership verification. Our approach is based on text-preserving watermarking using invisible Unicode sequences embedded into documents prior to fine-tuning. Each watermark is split into a \cue and a corresponding \sign, embedded in disjoint chunks of text. At audit time, targeted black-box challenges present the \cue; reproduction of the \sign in the model’s output indicates memorization consistent with training on the marked documents. Detection is formalized as a ranking-based hypothesis test against a reserved set of random counterfactual watermarks, yielding explicit and provable bound on the false-positive rate (FPR).

We evaluate our method across multiple open-source LLMs and text domains, including news and poetry. Experiments examine memorization, sensitivity to fine-tuning set size, multi-watermarks interference, true positive rate (TPR) and FPR. 
Results show multi-watermark support and reliable detection with no false positives observed in our experiments, and strong detection rates even when the marked documents constitute a small fraction of the fine-tuning data.

Our main contributions are:\\
(1) A novel, minimally invasive, text-preserving watermarking framework for post-hoc auditing of LLM fine-tuning data, based on invisible canaries and black-box challenges. \\
(2) A statistically grounded decision procedure using ranking against reserved counterfactual watermarks, providing provable control of the FPR and supporting large-scale, multi-user attribution. \\(3) An extensive empirical evaluation across models and text domains demonstrating robustness to scale, high TPR, low FPR, and resilience to multi-watermark interference.

We release code, watermark embedding and detection utilities, and evaluation scripts to facilitate reproducibility\footnote{All code and evaluation scripts are available at \url{https://github.com/liam-0/Data-Provenance-Auditing-of-Fine-Tuned-Large-Language-Models-with-a-Text-Preserving-Technique}.}

The remainder of the paper is organized as follows. Section~\ref{sec:pb} formulates the dataset provenance auditing problem and its requirements. Section~\ref{sec:proposal} describes our approach, including watermark design, embedding, detection, and FPR-bounded decision. Section~\ref{sec:experiments} presents experimental results, Section~\ref{sec:rw} reviews related work, Section~\ref{sec:discussion-limitations} discusses limitations, and Section~\ref{sec:ccl} concludes.

\section{Problem statement} 
\label{sec:pb}

\noindent\textbf{Dataset provenance auditing (DPA) problem.} We consider the problem of auditing whether a collection of high-quality documents has been used to fine-tune a LLM under the assumption that the documents can be proactively marked before potential usage. The goal is to enable reliable post-hoc detection of unauthorized usage.

We decompose such auditing procedures in two components: (1) a \emph{marking scheme} $mark$, which embeds imperceptible marks in documents prior to release, and (2) a \emph{verification procedure} $verif$, that queries the suspicious model and makes a decision from the resulting responses. 

Let $\col$ denote a protected dataset (e.g., a collection of news articles, poems...), marked prior to release. Let $\mathcal{L}$ denote a suspicious model fine-tuned on a dataset that may include $mark(\col)$. The verification procedure has access to both $mark(\col)$ and $\mathcal{L}$ and outputs a binary decision. Specifically, $verif(mark(\col),\mathcal{L}) \in \{0,1\}$, where output $1$ indicates the claim that $\col$ (or part of it) was used in the fine-tuning of $\mathcal{L}$, and output $0$ indicates no such claim. A valid DPA scheme must satisfy requirements at two levels: 

\noindent\textbf{Threat model.} We consider black-box dataset provenance auditing for fine-tuned LLMs. A data owner proactively marks documents before release and later audits a suspicious fine-tuned model through API access only. The verifier has access to the marked collection and model outputs, but not to model weights, gradients, logits, token probabilities, or training data. We assume the model provider may apply standard preprocessing or normalization during fine-tuning, but does not employ adaptive defenses explicitly designed to subvert provenance auditing. Attacks such as watermark forgery or targeted watermark removal are therefore considered out of scope and discussed as limitations.

\noindent\textbf{Data marking requirements.} A marking scheme $mark$ embeds marks into documents $\doc \in \col$ before possible fine-tuning and must satisfy the following: 

    \textit{Text authenticity.} The marking scheme must preserve the visible text exactly, ensuring that the author’s original wording and formatting remain unchanged.
    
\noindent\textit{Scalability.} The marking scheme must support the assignment of many distinct watermarks across users and documents, enabling multi-watermark attribution without collisions or interference. 

\noindent\textit{Robustness.} Marks must remain detectable despite common data processing and normalization pipelines; robustness is required only against non-adversarial transformation and filters. We discuss robustness to adversarial transformations with comparison to the existing works in limitations and in Appendix~\ref{app:robustness-transformations}.

\noindent\textbf{Verification procedure requirements.} Given a suspicious model $\mathcal{L}$ and a set of marked sensitive documents $mark(\col)$, the verification procedure $verif$ must satisfy:

    \textit{Black-box.} The procedure must operate using only input–output access to $\mathcal{L}$, without reliance on internal model parameters, gradients, or token probabilities. This reflects realistic deployment settings for proprietary or hosted LLMs. 
    
    \textit{Soundness.} Any positive claim $verif(mark(\col),\mathcal{L}) = 1$ must be correct except with a provably small probability. The FPR must be strictly bounded, so that positive findings constitute credible evidence of data usage. 
    
    \textit{Completeness.} If a sufficiently large number of marked documents (e.g., tens of documents) was used during fine-tuning, the procedure should detect it with high probability, ensuring that usage cannot systematically evade detection.

\section{Provenance Auditing through Non-Visible Watermarks} \label{sec:proposal}

\begin{figure*}[]
    \centering
    \hspace*{0.05\textwidth}
    \includegraphics[clip, trim=0cm 5cm 0.5cm 0cm, width=0.8\textwidth]{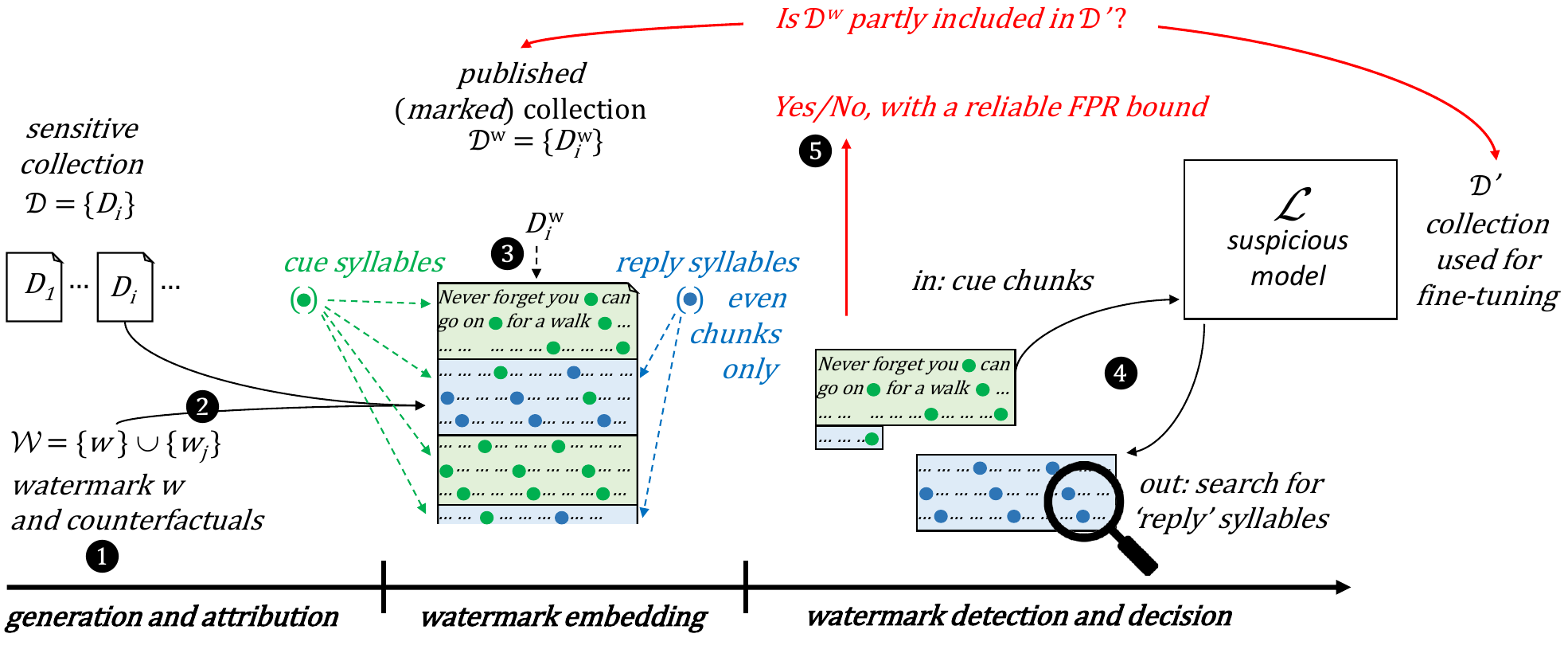}
    \caption{Overview of the proposal. A user is attributed a set $\ws_\lo$ of \lo watermarks and commits to one, \w, which is embedded into their sensitive collection prior to release. At audit time, the model is queried using the \cue of each watermark of $\ws_\lo$, which may elicit the corresponding \sign.  The target watermark \w is then ranked against counterfactuals. If it exceeds a fixed threshold $k$, we infer that sensitive data was included in the model’s fine-tuning set.}
    \label{fig:overview}
\end{figure*}

Our proposal is sketched in Figure~\ref{fig:overview} and described below. Watermarks are generated by pairing a \cue and a \sign, both composed of syllables of invisible characters (Section~\ref{sec:generation-WM}). When a user requests a watermark, they are attributed a set of \lo watermarks (Section~\ref{sec:attribution-WM}), commit to one watermark \w, and embed it into their sensitive collection prior to release (Section~\ref{sec:embedding-WM}). 

At audit time, a suspicious model is queried via black-box access using (green) chunks containing a \cue, and the resulting outputs are analyzed to detect the presence of the corresponding \sign (Section~\ref{sec:detection-WM}). This process is repeated using the \lo-1 left-out watermarks, which act as counterfactuals and enable statistically grounded decision: watermarks are ranked according to the number of times their \sign is detected. If the publication watermark ranks above a fixed threshold~$k$, we conclude that sensitive data was included in the model’s fine-tuning set (Section~\ref{sec:ranking}), with false positive rate at most~$k/\lo$. Detailed algorithms and analyses are provided in Appendix~\ref{algs}.

\subsection{Watermark generation} \label{sec:generation-WM}

Our watermarks act as canaries, i.e., synthetic sequences designed to be
memorized during training and selectively reproduced at inference time. They are constructed from an \textit{alphabet} $\mathcal{A}$ composed of non-rendering characters with no impact on formatting, ensuring text-preservation. 
In preliminary experiments, isolated invisible characters yielded weak
detection. We therefore group invisible characters into short clusters, which we refer to as \textit{syllables}.
A syllable is an ordered list of \sSy characters of $\mathcal{A}$, denoted $\syllable \in \mathcal{A}^\sSy$. 
A \textit{watermark} $\w$ is an ordered list of \sW syllables,
$\w = (\syllable_i)_{i = 1 ..\sW}$, $\syllable \in \mathcal{A}^\sSy$.

Each watermark consists of a \cue and a \sign, denoted $\w c$ and $\w r$ respectively, with:
$\w c = (\syllable_i)_{\syllable_i \in \w, i \leq \sCu}$
and $\w r = (\syllable_i)_{\syllable_i \in \w, i > \sCu}$

We call the \textit{tail of a \cue} its last $\cinr$ syllables. The \cue is embedded in a prompt and serves as the stimulus for the appearance of the \sign. 

We note $\ws$ the domain of watermarks. To minimize false positives, we aim to reduce the likelihood that a \sign associated with a dataset is observed by a model not trained on it. To this end, we impose the following constraints on $\ws$:

(1) Each \cue corresponds to a unique \sign and vice versa: $\forall (\w_1, \w_2) \in \ws^2,\;
    \bigl(\w c_1=\w c_2 \vee \w r_1=\w r_2\bigr)
    \Leftrightarrow \w_1 = \w_2$
    
(2) No \cue is contained in a \sign and vice versa:
$\forall (\w_1, \w_2) \in \ws^2,$ $\neg\bigl(\w c_1 \sqsubseteq (\w r_2)\bigr) \land\;
\neg\bigl(\w r_1 \sqsubseteq (\w c_2)\bigr)$
where $x \sqsubseteq y$ denote that the ordered list 
$x$ is a contiguous subsequence of the ordered list 
$y$.
 
Assuming $\sCu \geq \sW - \sCu$, we have (see Appendix~\ref{algs}): 
\[
\begin{aligned}
 |\ws| \geq & \max_{x \in [1,|\mathcal{A}|^{\sSy(\sW-\sCu)}]} \\ & \min(  x,  |\mathcal{A}|^{\sSy\sCu}-x\times(\sSy(2\sCu - \sW) +1)  |\mathcal{A}|^{\sSy(2\sCu - \sW)})
 \end{aligned}
\]

In particular, with $\sW = 2*\sCu$,  $|\ws| \geq \frac{|\mathcal{A}|^{\sSy\sCu}}{2}$.

\subsection{Watermark Attribution} \label{sec:attribution-WM}

We assume the presence of a trusted entity responsible for generating and assigning watermarks while enforcing the constraints defined in Sec.~\ref{sec:generation-WM}. This entity guarantees the uniqueness of each watermark by maintaining a record of all previously communicated watermarks, denoted $\ws_{com}$.

Whenever a user requests a new watermark, the trusted entity samples a set $\ws_\lo$ uniformly at random from $\ws \setminus (\ws_{com})$, updates its record $\ws_{com} \leftarrow{}\ws_{com} \cup \ws_\lo$, and returns $\ws_\lo$. 

$\ws_\lo$ serves as a candidate set of watermarks for the user: they select one uniformly at random from $\ws_\lo$ for actual use and  retain the remaining watermarks as counterfactuals for the decision procedure (see Sec.~\ref{sec:ranking}). 

\subsection{Watermark Embedding} \label{sec:embedding-WM}

A document $\doc$ is first segmented into an ordered list of text units, $(\word_i)_{i=1..\Delta}$. In whitespace-delimited languages, these units can be obtained by splitting on spaces; in languages without explicit word boundaries, they can be obtained using standard language-specific segmentation tools.

Because the \cue is used to construct prompts in which no \sign should appear, the two are embedded in separate chunks of the document. However, to encourage the model to associate the tail of the \cue with the corresponding \sign, the last $\cinr$ syllable of the \cue to be embedded alongside the \sign. 

To ensure that 1) significant samples of the document necessarily contains both the \cue and the \sign, and 2) that \cue chunks fit within context window, we chunk \doc into $\lceil\Delta/\delta\rceil$ sub-documents ${(\chunk)}_{i=1..\lceil\Delta/\delta\rceil}$.
 Sub-documents with odd indexes, or \cue chunks, are embedded with the first $\sCu-\cinr$ syllables of the \cue, i.e. the \cue minus its tail. Sub-documents with even indexes are called \sign chunks and embedded with $(\syllable_i)_{s_i \in w, i > \sCu-\cinr}$, i.e., the tail of the \cue plus the \sign.  Syllables are inserted cyclically every \spacing word. 
 
The resulting watermarked document consists of an ordered sequence of syllables (each containing \sSy invisible characters) interleaved with the original words, which remain unchanged and in their original order. Appendix~\ref{app:embed} details the embedding algorithm and the number of watermark repetitions it inserts in a document.

\subsection{Watermark Detection} \label{sec:detection-WM}

\noindent\textbf{Single document.} For a document $\doc$ watermarked with $\w$, denoted $\doc^\w$, verification proceeds by constructing
$\lfloor \tfrac{\lceil \Delta / \delta \rceil}{2} \rfloor$ prompts $ID^\w_i$.
Each prompt consists of a $\cue$ chunk concatenated with the beginning of the subsequent $\sign$ chunk, truncated after its first $\cinr$ invisible syllables so as to complete the $\cue$.

The chunk-level verification procedure, described in Algorithm~\ref{alg:challenge}, consists in prompting $\mathcal{L}$ $\lambda$ times with $ID^\w_i$. It outputs 1 if and only if the sign $\w r$ is detected in the output of $\mathcal{L}$ at least once. The document-level verification score of a document $\doc^\w$, is the sum of its chunk-level scores:
\[
    verif(\mathcal{L}, \doc^\w, \lambda) = \sum_{i=1..\lfloor\frac{\lceil \Delta/\delta \rceil}{2}\rfloor} verif(\mathcal{L}, ID^\w_i,\lambda)
\]

\noindent\textbf{Collection of documents.}  The score of an entire collection of documents 
\(\col\) embedded with the same watermark \(\w\), denoted \(\col^\w\),  is the sum of 
each document's score:
\[
    verif(\mathcal{L}, \col^\w, \lambda) = \sum_{\doc^\w \in \col^\w} verif(\mathcal{L}, \doc^\w,\lambda)
\]

\subsection{Decision process}
\label{sec:ranking}

Given a watermarked collection $\col^\w$, we compare its score given by \textit{verif($\mathcal{L}, \col^\w, \lambda$)} to the score of counterfactuals, i.e. watermarks that were never used and cannot have been used in fine-tuning. Since $\w$ is randomly sampled from a candidate set $\ws_\lo$ , left-out elements may serve as counterfactuals. Importantly, a watermark cannot be scored \textit{ex nihilo}: 
the scores of the counterfactual watermarks must be computed in the same context as the one being tested.  Thus, \textit{verif($\mathcal{L}, \col^\w, \lambda$)} is ranked compared to \textit{verif($\mathcal{L}, \col^{w_\lo}, \lambda$)} for each $\w_{\lo} \in \ws_{\lo} \backslash \{\w\}$. Membership is claimed for a watermark if its score is $>0$ and its rank $\geq k$ for a given $k$. The procedure is given in Algorithm~\ref{alg:test}.

While not providing an auditing approach, \citet{10992516} discussed the properties they should verify to obtain sound and tractable membership proofs. In particular, they demonstrated that the FPR of the ranking test is bounded if: (1) target data ($\w$) is explicitly sampled uniformly at random from a candidate set ($\ws_\lo$) whose left-out members serve as counterfactuals; (2) the injected data does not impact other data published afterwards, e.g. if it carries no useful information. Since our proposed watermarks are invisible, they are unlikely to impact the production of subsequent documents by third parties. Hence, our approach satisfies these conditions and, in line with~\cite{10992516}, we have FPR $\leq \frac{k}{\lo}$.

\subsection{Practical Considerations} \label{sec:conclu-proposal}

Certain requirements of Section~\ref{sec:pb} are achieved by design: \textbf{text authenticity} follows from the use of non-rendering Unicode which leaves the visible text sequence intact; 
\textbf{black-box operability} stems from the verification requiring text I/Os only; \textbf{scalability} stems from the exponential growth of the watermark space; and \textbf{soundness} is ensured by the ranking test. In practice, we record a cryptographic commitment to $\w$ at assignment time, which prevents post-hoc cherry-picking. Beyond these design properties, several practical aspects must be assessed empirically:

\noindent\textbf{Choice for the threshold $k$.} The ranking test gives a simple theoretical rule ($\text{FPR} \leq k/\lo$). The rate at which left-out watermarks are regurgitated must be verified in practice.

\noindent\textbf{Completeness and verification budget (in I/Os).} Completeness depends on the per-chunk \emph{hit} probability $p$ of a single chunk-level verification success. 
Repeating the prompt $\lambda$ times reduces the per-chunk \emph{miss} probability to $(1-p)^\lambda$. Probing $X$ independent chunks from the same collection, the collection-level probability of having no \textit{hit} is $p_{FN} = (1-p)^{\lambda X}$. Estimating realistic values of $p$ across models and domains and instantiating $(\lambda,X)$ to achieve practical error budgets, requires empirical measurement.

\noindent\textbf{Scalability in practice.} While the watermark space $|\ws|$ grows exponentially in theory, its effective use must be quantified, including collision risk and attribution accuracy under multi-watermark attribution.

\noindent\textbf{Robustness.} Appendix~\ref{app:rob} emprically demonstrates that watermarks are robust to common non-adversarial transformations they: (1) survive to insertion and recovery from 4 websites known to be scrapped for data collection such as Wikipedia and GitHub; (2) withstand 6 standard data cleaning and processing pipelines, including C4~\cite{10.5555/3455716.3455856}, CCNet~\cite{wenzek-etal-2020-ccnet}, the Pile~\cite{gao2020pile800gbdatasetdiverse}, RedPajama \cite{10.5555/3737916.3741613}, Dolma~\cite{soldaini-etal-2024-dolma} and FineWeb~\cite{10.5555/3737916.3738886}; (3) are reliably recognized by 10 representative tokenizers (including tokenizers used by Mistral-7B-v0.1, DeepSeek-R1 and gpt-4o); and (4) are not filtered by safety mechanisms in public chatbot interfaces like ChatGPT, Le Chat, and DeepSeek.

\section{Experimental Evaluation} \label{sec:experiments}

This section presents the experimental evaluation of our proposed approach. We focus on its detection power, measured as the average regurgitation rate of a \sign under varying conditions. Specifically, we investigate how regurgitation depends on (i) the size of the watermarked collection, (ii) the total fine-tuning dataset size, and (iii) the number of unique watermarks. We further show that spurious regurgitation is negligible, yielding high TPR@0\%FPR even in challenging settings where the closest existing baseline fails.

\subsection{Experimental settings}

\noindent\textbf{Datasets.} 
We evaluate our method on three datasets representative of personal and copyrighted textual content. 
(1) \href{https://zenodo.org/records/7455623}{\textbf{Blog1k}}, a subset of the Blog Authorship Corpus, containing\textit{personal blog} posts. 
(2) \href{https://www.kaggle.com/datasets/tgdivy/poetry-foundation-poems}{\textbf{Poems}}, containing \textit{poems} collected from the Poetry Foundation. 
(3) \href{https://huggingface.co/datasets/abisee/cnn_dailymail}{\textbf{News}} (CNN/DailyMail) containng longer news articles, enabling evaluation in long-context generation settings. Entries shorter than 200 words are discarded, reducing Blog1k to 17,843 documents with an average length of 360.03 words, Poems to 5,636 documents of 480.01 words on average, and News to 11,837 documents averaging 1,519.96 words.

\noindent\textbf{LLMs $\mathcal{L}$.} All experiments are conducted using LLaMA-2-7b-hf and Mistral-7B-v0.1 (denoted as LLaMA and Mistral, respectively), two prominent open-source models, ensuring reproducibility. We limit each output to 200 tokens and detail the used parameters in Appendix~\ref{sec:xpdetails}.

\noindent\textbf{Parametrization of the approach.}
The number \sSy ~of characters per syllable entails a trade-off. If \sSy is too small, a syllable may be disregarded as noise; if it is too large, its final part may cease to be associated to preceding “legitimate” words, thereby diminishing regurgitation rates. Based on empirical observations, we set \sSy= 4. 

Similarly, $\sW$, the size of the watermark, must be appropriately calibrated to balance the trade-off between legitimate and spurious detection. If $\sW$ is too small, the watermark is more likely to occur “by chance”, whereas if it is too large, even minor deviations may prevent successful detection. Based on empirical evidence, we set $\sW$=8. We choose \sCu=5 and  \cinr=1 so that four syllables are embedded repeatedly in each document chunk, irrespective of parity.

The alphabet comprises $121$ invisible Unicode characters, yielding $|\ws| \geq 10^{24}$. The selection process is described in Appendix~\ref{app:alphabet}.

For the Blog1k and Poems datasets, given a document $\doc$ of length $\Delta$, we set $\delta = \lceil \Delta / 2 \rceil$ and partition $\doc$ into two consecutive chunks.
For News, we set $\delta = 200$, each $\doc$ being partitioned in approximately 8 chunks.

\noindent\textbf{Evaluation and default values.} We report the probability of successful chunk-level verification at $\lambda =1$ with 95\% confidence intervals. For each chunk, regurgitation is averaged over five watermarks and three prompts each, and the final probability is obtained by averaging over all chunks.

Unless otherwise stated, 1) datasets are downsampled to 1{,}000 entries for fine-tuning, with randomly selected subsets watermarked to form the sensitive collections; 2) the number of watermarked documents is 40 across all models and datasets, except for Mistral on Blog1k (60), following subsection~\ref{sec: Impact of the size of the watermarked collection} to target intermediate regurgitation regimes. Appendix~\ref{sec:stats} reports detailed dataset statistics.

\subsection{Regurgitation scales with the size of the watermarked collection}
\label{sec: Impact of the size of the watermarked collection}

\begin{figure}[]
    \centering
    \includegraphics[width=0.85\linewidth]{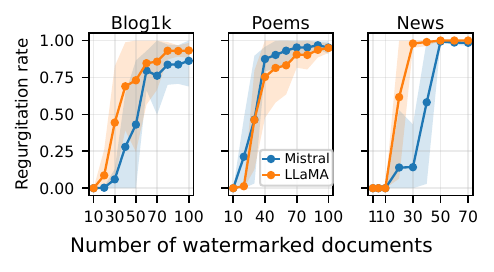}
    \caption{Average regurgitation rate as a function of watermarked collection size. Shaded areas indicate 95\% confidence intervals.}
    
    \label{fig:xp1_ci95}
\end{figure}

To evaluate how regurgitation scales with the size of the watermarked collection, we embed a single watermark \w into a subset $\col^\w$ of the fine-tuning dataset. For Blog1k and Poems, 
$|\col^\w|$ ranges from 10 to 100 documents (1-10\%), and for News, which contains longer documents split into more chunks, from 1 to 70 documents (0.1-7\%). Figure~\ref{fig:xp1_ci95} reports the chunk-level regurgitation rates.

\begin{table*}
\centering
\caption{Empirical estimation of chunk hit probability $p$ and collection miss probability $p_{FN}$.}
\label{tab:success-residual}
\begin{subtable}{0.6\columnwidth}
\centering
\caption{Blog1k}
\resizebox{\columnwidth}{!}{
\begin{tabular}{c|cc|cc}
\hline
\multirow{2}{*}{nbdocs} & \multicolumn{2}{c|}{Mistral} & \multicolumn{2}{c}{LLaMA} \\
 & $p$ & $p_{FN}$ & $p$ & $p_{FN}$ \\
\hline
30  & 0.060 & 0.156 & 0.444 & 2.2$\times 10^{-8}$ \\
40  & 0.280 & 2$\times 10^{-6}$ & 0.690 & 4.5$\times 10^{-21}$ \\
50  & 0.431 & 5.9$\times 10^{-13}$ & 0.731 & 3.3$\times 10^{-29}$ \\
\hline
\end{tabular}
}
\end{subtable}
\hfill 
\begin{subtable}{0.6\columnwidth}
\centering
\caption{Poems}
\resizebox{\columnwidth}{!}{
\begin{tabular}{c|cc|cc}
\hline
\multirow{2}{*}{nbdocs} & \multicolumn{2}{c|}{Mistral} & \multicolumn{2}{c}{LLaMA} \\
 & $p$ & $p_{FN}$ & $p$ & $p_{FN}$ \\
\hline
30  & 0.464 & 7.3$\times 10^{-9}$ & 0.462 & 8.3$\times 10^{-9}$ \\
40  & 0.875 & 7.5$\times 10^{-37}$ & 0.753 & 4.8$\times 10^{-25}$ \\
50  & 0.903 & 2.6$\times 10^{-51}$ & 0.815 & 2.5$\times 10^{-37}$ \\
\hline
\end{tabular}
}
\end{subtable}
\hfill 
\begin{subtable}{0.6\columnwidth}
\centering
\caption{News}
\resizebox{\columnwidth}{!}{
\begin{tabular}{c|cc|cc}
\hline
\multirow{2}{*}{\makecell{nbdocs\\ }} & \multicolumn{2}{c|}{Mistral} & \multicolumn{2}{c}{LLaMA} \\
 & $p$ & $p_{FN}$ & $p$ & $p_{FN}$ \\
\hline
30 & 0.039 & 8.6$\times 10^{-3}$  & 0.677 & $\approx 10^{-59}$ \\
40 & 0.298 & $\approx 10^{-25}$   & 0.815 & $\approx 10^{-117}$ \\
50 & 0.772 & $\approx 10^{-129}$  & 0.898 & $\approx 10^{-199}$ \\
\hline
\end{tabular}
}
\end{subtable}
\end{table*}

Regurgitation increases monotonically with $|\col^\w|$. For small watermarked subsets, the rate remains near zero; beyond a model- and dataset-dependent threshold, it increases sharply before saturating between 0.75 and 0.98. For instance, on Blog1k, this transition occurs at 
$|\col^\w|\approx40$ for LLaMA and $|\col^\w| \approx 60$ for Mistral. Despite these differences, both models converge to a similar saturation level of approximately 0.75 by $|\col^\w|=70$. Table~\ref{tab:success-residual} summarizes the corresponding hit and miss probabilities at the collection level.

\noindent \begin{leftbar}
\textbf{Completeness.} 
This experiment provides an empirical estimate of the hit probability $p$ for a single prompt and the miss probability $p_{FN}$ over a collection. For both models and all datasets, $p_{FN}<10^{-2}$ with 30 watermarked documents, decreasing to as low as $10^{-59}$, confirming that watermarks remain highly detectable even at moderate collection sizes.
\end{leftbar}

\subsection{Regurgitation is not systematically reduced by increasing the number of unmarked documents}
\label{sec: Impact of the size of the dataset}
\begin{figure}[]
    \centering
    \includegraphics[width=0.85\linewidth]{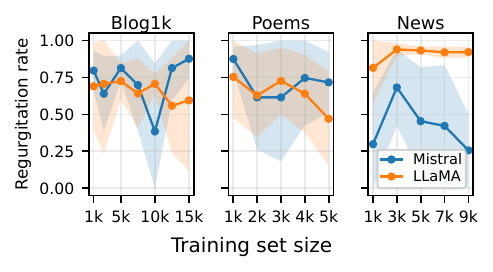}
    \caption{Average regurgitation rate as a function of fine-tuning dataset size. Shaded areas denote 95\% confidence intervals.}
    \label{fig:xp2_ci95}
\end{figure}

To study the effect of fine-tuning dataset size, we fix $|\col^\w|$ and vary the amount of unmarked data. Across settings, the proportion of watermarked documents ranges from a maximum of 6\% down to a minimum of 0.25\% of the dataset. Figure~\ref{fig:xp2_ci95} reports chunk-level regurgitation rates.

 Across all datasets and models, increasing the amount of unmarked document does not produce a monotonic reduction in regurgitation. Instead, rates fluctuates within a relatively stable range.
On Blog1k and Poems, both models exhibit non-monotonic behavior as the fine-tuning set grows, with Mistral showing higher variability. On News, LLaMA maintains consistently high regurgitation across all training scales, whereas Mistral shows pronounced fluctuations, peaking at 0.74 with 3{,}000 fine-tuning documents, compared to approximately 0.25 at 1{,}000 and 9{,}000.

We hypothesize that this variability primarily stems from the relatively small fine-tuning datasets, which may amplify sensitivity to stochastic effects such as data sampling and optimization noise. Consequently, the effect of fine-tuning set size is difficult to disentangle from this variability.

\noindent \begin{leftbar}
\textbf{Scalability w.r.t. finetuning dataset}.
Increasing the number of unmarked documents does not yield predictable changes in regurgitation, which remains largely non-monotonic. The fraction of watermarked documents alone is a poor predictor of regurgitation strength.
\end{leftbar}

\subsection{Verification support multi-watermark settings}
\label{sec: Impact of the number of unique watermarks}
\begin{figure}[htbp]
    \centering
    \includegraphics[width=0.85\linewidth]{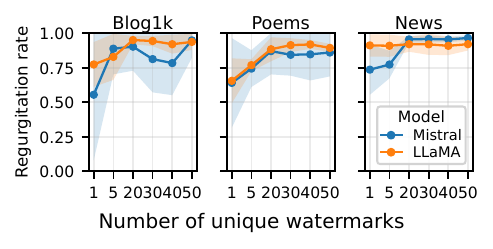}
    \caption{Average regurgitation rate as a function of the number of unique watermarks in the fine-tuning dataset. Shaded areas denote 95\% confidence intervals.}
    \label{fig:xp3_ci95}
\end{figure}

We investigate the impact of the number of distinct watermarks present in the fine-tuning dataset, denoted $U$, on the verification procedure. Unlike previous experiments, where the five watermarks were embedded in the same collection and evaluated independently, each watermark is embedded into a disjoint subset of the dataset. For $U=1$, watermarks are evaluated in isolation, and are all present in the fine-tuning dataset starting at 
$U=5$, set to 5,000 documents to accommodate up to $50$ unique watermarks. Figure~\ref{fig:xp3_ci95} shows the aggregated chunk-level regurgitation rate.  Detailed per-watermark results are provided in Appendix~\ref{app: Details of unique watermarks experiment}.

 As $U$ increases from $1$ to moderate values ($U=5$ and $U=20$), regurgitation rises across all datasets before stabilizing near saturation for both models (approximately $0.85$–$0.95$ on Blog1k and Poems, and above $0.9$ on News).  This trend is particularly pronounced for Mistral, whose regurgitation increases from roughly $0.5$–$0.7$ at $U=1$ to above $0.85$ at $U=50$ across all datasets, indicating that exposure to multiple distinct watermarks does not diminish, and may even strengthen, memorization.

To examine whether this behavior generalizes to more recent and larger-scale models, we extend the analysis to GPT-OSS-20B (Appendix~\ref{app: Details of unique watermarks experiment}) and observe the same trend. Fine-tuning parameters and watermarked-text statistics are reported in Appendices~\ref{sec:xpdetails} and~\ref{sec:stats}.

\noindent
\begin{leftbar}
\textbf{Scalability: multi-watermark support.}
Across all datasets and models, increasing the number of distinct watermarks does not reduce regurgitation rates; detection performance remains stable or improves under multi-watermark fine-tuning.
\end{leftbar}

\subsection{Decision yields high TPR at low FPR.}

While previous experiments focused on regurgitation rates, we now evaluate the \textit{decision procedure} for assessing membership following Algorithm~\ref{alg:test}. Let $\w_{i \in [1,5]}$ be the 5 watermarks used throughout the experiments, and let $\bar{\w}_{j, j \in [1,99]}$ be randomly generated counterfactual watermarks. For each dataset, each $w_i$ is scored using $verif(\mathcal{L}_i,\col^{\w_i}, 1)$, where $\mathcal{L}_i$ denotes the model $\mathcal{L}$ fine-tuned on the dataset containing $\mathcal{D}^{\w_i}$. Counterfactual watermarks are evaluated via $verif(\mathcal{L}_i,\col^{\bar{\w}_j}, 1)$ with $i \in [1,5]$ selected at random. Table~\ref{tab:ranking} reports the resulting scores over three repetitions.

\begin{table}[h]
\centering
\caption{Average scores over three ranking tests. Legitimate watermarks ($\w_i$) yield substantial regurgitation, while counterfactuals ($\bar{\w}_j$) produce none.}
\resizebox{\linewidth}{!}{
\begin{tabular}{lc|ccc|c}
\toprule
Model Fine & Dataset   
& \multicolumn{3}{c}{Scores on $\w_{i, i \in [1,5]}$} 
& \multicolumn{1}{c}{All $\bar{\w}_j$}  \\
tuned on $\col^{\w_i}$ & ($|\col^{\w_i}|$) 
& Avg. & Worst $\w_i$ & Best $\w_i$ 
& \multicolumn{1}{c}{$j \in [1,99]$} \\
\midrule
\multirow{3}{*}{Mistral-7B} 
 & Blog1k (60)  & 48 & 39 & 57 & 0  \\
 & Poems (40)   & 35.2 & 32 & 38 & 0  \\
 & News  (40)   & 51.9 & 0 & 110.72 & 0  \\
\midrule
\multirow{3}{*}{LLaMA-2-7B} 
 & Blog1k (40)  & 27.6 & 16 & 35.2 & 0  \\
 & Poems (40)   & 30 & 0.37 & 36 & 0  \\
 & News (40)   & 141.86 & 86.5 & 169.54 & 0  \\
\bottomrule
\end{tabular}
}
\label{tab:ranking}
\vskip -0.1in
\end{table}

No spurious regurgitation (i.e. regurgitation of the \sign of a counterfactual watermark) is detected over more than $10^5$ challenges. In contrast, legitimate watermarks generally exhibit substantial regurgitation, with average rates above 0.5. A single exception occurs for  Mistral fine-tuned on News, where one watermark yields 0 regurgitation. Overall, these results correspond to an approximate TPR of 96.7\% at 0\% FPR with k=1.

\noindent \begin{leftbar}
\textbf{96.7\%@0\%FPR}. Spurious detections are negligible, suggesting that a small k=1 can be safely adopted. Additionally, verification may be performed on a subset of the sensitive collection to reduce I/O costs without compromising detection.
\end{leftbar}

\subsection{Closest existing work fails under multi-watermark evaluation}
To the best of our knowledge, the homoglyph watermarking scheme of~\cite{wei-etal-2024-proving} is the only existing text-preserving watermarking approach (see Section~\ref{sec:rw}). We evaluate this baseline under the setting of Section~\ref{sec: Impact of the number of unique watermarks} comprising 50 unique watermarks.

\begin{figure}[htbp]
    \centering
    \includegraphics[width=0.85\linewidth]{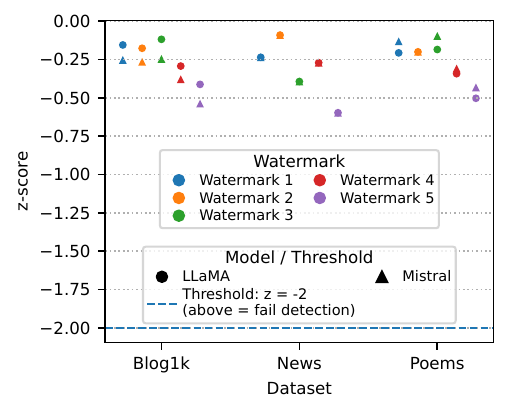}
    \caption{
Z-scores for the homoglyph watermarking baseline~\cite{wei-etal-2024-proving}. Each point corresponds to one watermark. The dashed line indicates the significance threshold ($p\approx 0.05$) above which score fails to achieve statistically significant detection.
}

    \label{fig:baseline}
\end{figure}

The baseline relies on average token loss and empirical Z-scores computed from a null distribution of 100 random sequences. Figure~\ref{fig:baseline} shows the resulting Z-scores, which are concentrated around $Z \approx -0.1$ to $-0.6$ corresponding to statistically non-significant detection ($p>0.2$). This indicates that the baseline fails to reliably distinguish watermarked samples from the null distribution in this regime.

In contrast, our method produces reliable membership signals across all datasets and models, exhibiting regurgitation rates $>0.8$ under the same setting (see Section.~\ref{sec: Impact of the number of unique watermarks}).

\begin{leftbar}
\textbf{Baseline detection fails in challenging regimes.}
Across all datasets and models, the homoglyph watermarking scheme~\cite{wei-etal-2024-proving} fails to produce statistically significant detection under multi-watermark evaluation. In contrast, our method consistently yields strong and reliable membership signals.
\end{leftbar}

\section{Related work}
\label{sec:rw}

\begin{table}[t]
  \caption{DPA techniques for free text generation and their capacity to preserve the text (Auth.), operate under a black-box model (BB), and provide statistically grounded proofs (Sound).}
  \label{tab:rw}
  \centering
  \resizebox{\columnwidth}{!}{
  \scriptsize
  \begin{tabular}{
    >{\centering\arraybackslash}p{0.55\columnwidth}|
    >{\centering\arraybackslash}p{0.08\columnwidth}|
    >{\centering\arraybackslash}p{0.15\columnwidth}|
    >{\centering\arraybackslash}p{0.08\columnwidth}
  }
    \toprule
    Approaches  &  Auth.      & BB & Sound \\
    \midrule
    Non-intrusive MIAs, e.g.~\cite{10.5555/3692070.3692545,puerto-etal-2025-scaling}  
      & \textcolor{ForestGreen}{\textbf{$\surd$}} & Some & \textcolor{WildStrawberry}{\textbf{$\times$}} \\
    \textit{Waterfall}~\cite{lau2024waterfall} 
      & \textcolor{WildStrawberry}{\textbf{$\times$}} & \textcolor{ForestGreen}{\textbf{$\surd$}} & \textcolor{WildStrawberry}{\textbf{$\times$}} \\
    \textit{STAMP}~\cite{rastogistamp}   
      & \textcolor{WildStrawberry}{\textbf{$\times$}} & \textcolor{WildStrawberry}{\textbf{$\times$}} & \textcolor{ForestGreen}{\textbf{$\surd$}} \\
    Fictuous knowledge~\cite{cui-etal-2025-robust} 
      & \textcolor{WildStrawberry}{\textbf{$\times$}} & \textcolor{ForestGreen}{\textbf{$\surd$}} & \textcolor{ForestGreen}{\textbf{$\surd$}} \\
    Copyright traps~\cite{10.5555/3692070.3693506}
      & \textcolor{WildStrawberry}{\textbf{$\times$}} & \textcolor{WildStrawberry}{\textbf{$\times$}} & \textcolor{WildStrawberry}{\textbf{$\times$}} \\
    Hash~\cite{wei-etal-2024-proving} 
      & \textcolor{WildStrawberry}{\textbf{$\times$}} & \textcolor{WildStrawberry}{\textbf{$\times$}} & \textcolor{ForestGreen}{\textbf{$\surd$}} \\
     Homoglyphs~\cite{wei-etal-2024-proving} 
      & \textcolor{ForestGreen}{\textbf{$\surd$}} & \textcolor{WildStrawberry}{\textbf{$\times$}} & \textcolor{ForestGreen}{\textbf{$\surd$}} \\
    Ours 
      & \textcolor{ForestGreen}{\textbf{$\surd$}} & \textcolor{ForestGreen}{\textbf{$\surd$}} & \textcolor{ForestGreen}{\textbf{$\surd$}} \\
    \bottomrule
  \end{tabular}
  }
  \vskip -0.1in
\end{table}

\subsection{Generative AI, watermarking, and DPA}

Prior work on watermarking in generative AI largely assumes that the model provider controls the watermarking process~\cite{sokgenai}. Most approaches focus on watermarking model outputs to detect AI-generated content, to protect model copyright, where watermarks are used to trace content or derived models back to a specific source-model~\cite{
10.5555/3618408.3620182,10.5555/3600270.3600662,DBLP:conf/emnlp/ZhaoLW22}, or to associate generated text with sources in their training data~\cite{lu-etal-2025-wasa}. We focus on dataset watermarking for DPA, whose goal is to enable a data owner to detect and \emph{prove} unauthorized use of their data during model training or adaptation. 
As noted by a recent survey~\cite{du2025sok}, ``insufficient attention has been
paid to the dataset copyright issues posed by LLMs''. Existing DPA techniques predominantly focus on the image domain (e.g.,~\citealp{wenger2022data,chen2025anonymityunveiledpracticalframework}), or on classification tasks, where provenance is inferred from predicted labels (e.g.,~\citealp{pmlr-v139-choquette-choo21a,li2022untargeted,liu2024watermarkingtextdatalarge,10.1145/3606274.3606279}).

\subsection{DPA in LLMs for free text generation}
DPA methods are commonly categorized by their degree of intrusiveness~\cite{chen2025anonymityunveiledpracticalframework,du2025sok}.

\noindent\textbf{Non-intrusive DPA}
 methods aim to detect usage without modifying the original data, typically using decision-boundary analysis for classification task or model-based analysis~\cite{du2025sok}.
 The latter largely overlaps with MIAs~\cite{Shokri2017}, a form of privacy attack increasingly adapted to DPA~\cite{10992516}. However, the effectiveness of such approaches on free-text generation has been increasingly questioned~\cite{duan2024membership,DBLP:conf/satml/MeeusSJFRM25}. While some recent work show promising detection~\cite{10.5555/3692070.3692545} in particular at the dataset level~\cite{maini2024llm,puerto-etal-2025-scaling}, the difficulty of sampling from the null hypothesis questions their ability to provide sound training-data proofs~\cite{10992516}. 
 
\noindent\textbf{Intrusive DPA} techniques embeds detectable signals into the dataset. Recent work has leveraged the radioactivity of model output watermarking techniques~\cite{NEURIPS2024_2567c95f,zhao2025can} to adapt them for DPA~\cite{lau2024waterfall,rastogistamp}. Such approaches may introduce distortions or errors~\cite{lau2024waterfall,rastogistamp}.

Other approaches rely on the insertion of canaries, such as fictitious knowledge and incorrect facts~\cite{cui-etal-2025-robust} or LLM-generated ``copyright traps''~\cite{10.5555/3692070.3693506}. Such canaries are perceptible, and it may be unacceptable to insert incorrect knowledge or repeated long sequences in a text. Moreover, copyright traps~\cite{10.5555/3692070.3693506} do not offer statistically grounded guarantees~\cite{10992516,wei-etal-2024-proving}.

\citet{wei-etal-2024-proving} perturbs documents with hash sequences or Unicode homoglyph substitutions, enabling hypothesis testing to verify whether a collection was used in LLM pretraining. To the best of our knowledge, the homoglyph-based approach is the only text-preserving (although some homoglyphs remain visually distinguishable) DPA technique supporting statistically grounded proofs.  As with our method, it is not robust to adversarial transformations; however, unlike our approach, it requires word probabilities and does not support multiple concurrent watermarks.

\section{Discussion and Limitations} \label{sec:discussion-limitations}

While our framework provides sound guarantees of authenticity, black-box deployability, and provable false-positive control, several limitations and broader considerations remain, which could be considered future work.


\noindent\textbf{Robustness to active removal.}
Our threat model assumes non-adversarial preprocessing, excluding active watermark removal. As shown in Appendix~\ref{app:rob}, the proposed scheme is largely preserved under realistic data-collection, preprocessing, tokenization, and public-interface workflows, supporting use cases in which no adversarial removal step is applied. However, the method is inherently brittle to an adversary aware of the watermark, who can apply simple targeted removal strategies such as stripping zero-width characters or filtering suspicious Unicode patterns. This limitation is shared by prior text-watermarking schemes for provenance and copyright protection~\cite{du2025sok,wei-etal-2024-proving}.

We further evaluate robustness under untargeted, generic removal attacks in Appendix~\ref{app:robustness-transformations}, and find that the method is resilient to a broad class of low-cost, semantics-preserving perturbations, and substantially more robust than prior work. In contrast, it remains vulnerable to high-cost attacks that leverage external models to substantially rewrite the text prior to training.

Importantly, such attacks affect completeness rather than soundness: they may reduce detection power and increase false negatives, but they do not compromise the false-positive control of the ranking-based test. We leave systematic adversarial hardening to future work.

 \noindent\textbf{Robustness to forgery and replay attacks.}
The protocol mitigates ex-post watermark forgery through the candidate-set and commitment mechanism. The used watermark is sampled from a candidate
set before auditing, unused candidates serve as counterfactuals in the ranking test, and a commitment to the selected watermark is recorded at assignment time. This prevents constructing or cherry-picking a watermark after observing the suspicious model. A stronger contamination or replay attack remains possible if an adversary deliberately reuses a known watermark, or injects it into unrelated data. We emphasize that, as in the rest of this work, we do not aim to provide guarantees under an active adversary. This setting is outside the scope of the current protocol and remains an important direction for future work.

\noindent\textbf{Trusted third party (TTP) considerations.} We rely on a TTP to generate (\cue,\sign) pairs obeying the global uniqueness and overlap constraints, to allocate candidate sets uniformly, and to keep public commitments to protect against cherry-picking. This simplifies soundness and prevents accidental collisions across users, but it introduces trust and operational burdens. Certain directions could reduce that burden like resorting to append-only public logs. 

\noindent\textbf{Scope and training-strategy dependency.} Our method is designed for provenance in supervised fine-tuning, where the \cue$\xrightarrow{}$ \sign signal concentrates in adapted parameters and supports practical black-box detection. We therefore study adapter-based fine-tuning~\citep{hu2022lora} as a representative deployment setting. Other paradigms --such as large-scale pretraining, reinforcement learning from human feedback~\cite{NIPS2017_d5e2c0ad, NEURIPS2022_b1efde53}, or some full fine-tuning procedures-- may attenuate idiosyncratic signals and reduce detection power, and are not claimed to be covered. Exploring these regimes constitutes natural next steps.

\noindent\textbf{Legal and ethical considerations.} Proposed marks are imperceptible, but their insertion alters the raw byte representation of a text, which may be unacceptable in certain areas (e.g., for official electronic documents). Furthermore, the probative value of watermark-based audits in litigation will depend on the evolution of legal standards of evidence. On this point, the technical results may well encourage judges to reason by presumption in order to facilitate the proof of unauthorised use of content, particularly content protected by copyright. In any case, it would be appropriate for the technical audit to be supplemented by governance frameworks and explicit consent mechanisms.

\section{Conclusion} \label{sec:ccl}

We presented a minimally invasive, text-preserving framework for dataset provenance auditing of fine-tuned LLMs, embedding invisible Unicode canaries in a \cue–\sign structure and detecting them via black-box prompting. The decision is statistically grounded, with a formally bounded FPR. 

Across multiple LLMs and datasets, the approach reliably supports multi-watermark attribution and achieves high TPR@0\%FPR with only tens of watermarked documents constituting as little as 0.25\% of the total fine-tuning dataset. These results highlight the approach’s practicality for provenance tracking in realistic fine-tuning scenarios.

\section*{Author Contributions}

This work was carried out as a collaborative effort combining technical, methodological, legal, and security perspectives. Yanming Li contributed to the software development and experiments. Cédric Eichler contributed to the methodology, experimental validation, and writing. Nicolas Anciaux contributed to the conceptualization and scientific coordination of the project, as well as to the writing. Alexandra Bensamoun contributed to the legal analysis and to the framing of the topic. Lorena Gonzalez Manzano contributed to the security analysis. Seifeddine Ghozzi contributed to the preliminary experiments. All authors contributed to discussions and reviewed the manuscript.

\section*{Acknowledgements}

This research was supported by grants ANR-22-PECY-0002 (IPoP project), by the DATAIA Convergence Institute as part of the Investments for the Programme d'Investissement d'Avenir (ANR-17-CONV-0003) managed by Inria Saclay, and by PID2023-150310OB-I00 (MORE4AIO project) funded by MCIU, AEI and FEDER. We also thank Inria and U3CM for the \href{https://team.inria.fr/petscraft/petsai/}{PETSAI} associated team grant which supports collaborative work between \href{https://team.inria.fr/petscraft}{Petscraft} and \href{https://cosec.inf.uc3m.es/}{COSEC}.


\section*{Impact Statement}
Potential societal consequences of our work might be on the juridical side.
Indeed, the need for technological support is more evident than the law's inability to ensure legal certainty in a copyright protection context. While there is indeed an exception at European level for text and data mining that suspends the application of copyright and related rights, which could perhaps benefit model training, this is subject to the dual condition that the AI provider has lawful access to the content and that the rights holder has not exercised his opt-out right. However, the opacity surrounding these activities makes it impossible to know whether the conditions are being met.
Therefore, the AI Act\footnote{See the AI Act enacted in June 2024 (\href{https://eur-lex.europa.eu/legal-content/EN/TXT/HTML/?uri=OJ:L_202401689}{link}).} imposed a transparency obligation on AI providers, which consists of developing and making available to the public a ``sufficiently detailed summary'' of the content used, for which a model was provided by the AI Office\footnote{See the AI Office enacted in July 2025 (\href{https://digital-strategy.ec.europa.eu/en/library/explanatory-notice-and-template-public-summary-training-content-general-purpose-ai-models}{link}).}. However, it is not certain that the required level will enable rights holders to enforce their rights. 
The situation is no more satisfactory in the United States, where some 50 lawsuits are pending on this issue and where the fair use exception is very uncertain. As proof, Anthropic agreed to settle a lawsuit by offering to pay \$1.5 billion in compensation for the unauthorized use of 500,000 books\footnote{See for example the news article by Reuters (\href{https://www.reuters.com/sustainability/boards-policy-regulation/anthropic-agrees-pay-15-billion-settle-author-class-action-2025-09-05/?utm_source=chatgpt.com}{link}).}. These uncertainties illustrate that legal instruments alone are unlikely to provide sufficient protection or enforceability, reinforcing the need for complementary technical solutions, such as the one presented in this paper.

\bibliographystyle{icml2026}
\bibliography{sample}

\begin{thebibliography}{51}
\providecommand{\natexlab}[1]{#1}
\providecommand{\url}[1]{\texttt{#1}}
\expandafter\ifx\csname urlstyle\endcsname\relax
  \providecommand{\doi}[1]{doi: #1}\else
  \providecommand{\doi}{doi: \begingroup \urlstyle{rm}\Url}\fi

\bibitem[Black et~al.(2021)Black, Leo, Wang, Leahy, and
  Biderman]{black_2021_5297715}
Black, S., Leo, G., Wang, P., Leahy, C., and Biderman, S.
\newblock {GPT-Neo}: Large scale autoregressive language modeling with
  mesh-tensorflow, March 2021.
\newblock URL \url{https://doi.org/10.5281/zenodo.5297715}.

\bibitem[Brown et~al.(2020)Brown, Mann, Ryder, Subbiah, Kaplan, Dhariwal,
  Neelakantan, Shyam, Sastry, Askell, Agarwal, Herbert-Voss, Krueger, Henighan,
  Child, Ramesh, Ziegler, Wu, Winter, Hesse, Chen, Sigler, Litwin, Gray, Chess,
  Clark, Berner, McCandlish, Radford, Sutskever, and
  Amodei]{NEURIPS2020_1457c0d6}
Brown, T., Mann, B., Ryder, N., Subbiah, M., Kaplan, J.~D., Dhariwal, P.,
  Neelakantan, A., Shyam, P., Sastry, G., Askell, A., Agarwal, S.,
  Herbert-Voss, A., Krueger, G., Henighan, T., Child, R., Ramesh, A., Ziegler,
  D., Wu, J., Winter, C., Hesse, C., Chen, M., Sigler, E., Litwin, M., Gray,
  S., Chess, B., Clark, J., Berner, C., McCandlish, S., Radford, A., Sutskever,
  I., and Amodei, D.
\newblock Language models are few-shot learners.
\newblock In Larochelle, H., Ranzato, M., Hadsell, R., Balcan, M., and Lin, H.
  (eds.), \emph{Advances in Neural Information Processing Systems}, volume~33,
  pp.\  1877--1901. Curran Associates, Inc., 2020.
\newblock URL
  \url{https://proceedings.neurips.cc/paper_files/paper/2020/file/1457c0d6bfcb4967418bfb8ac142f64a-Paper.pdf}.

\bibitem[Carlini et~al.(2023)Carlini, Ippolito, Jagielski, Lee, Tramer, and
  Zhang]{carlini2022quantifying}
Carlini, N., Ippolito, D., Jagielski, M., Lee, K., Tramer, F., and Zhang, C.
\newblock Quantifying memorization across neural language models.
\newblock In \emph{The Eleventh International Conference on Learning
  Representations}, 2023.
\newblock URL \url{https://openreview.net/forum?id=TatRHT_1cK}.

\bibitem[Chen \& Pattabiraman(2025)Chen and
  Pattabiraman]{chen2025anonymityunveiledpracticalframework}
Chen, Z. and Pattabiraman, K.
\newblock Anonymity unveiled: A practical framework for auditing data use in
  deep learning models.
\newblock In \emph{Proceedings of the 2025 ACM SIGSAC Conference on Computer
  and Communications Security}, CCS '25, pp.\  513–527, New York, NY, USA,
  2025. Association for Computing Machinery.
\newblock ISBN 9798400715259.
\newblock \doi{10.1145/3719027.3744794}.
\newblock URL \url{https://doi.org/10.1145/3719027.3744794}.

\bibitem[Choquette-Choo et~al.(2021)Choquette-Choo, Tramer, Carlini, and
  Papernot]{pmlr-v139-choquette-choo21a}
Choquette-Choo, C.~A., Tramer, F., Carlini, N., and Papernot, N.
\newblock Label-only membership inference attacks.
\newblock In Meila, M. and Zhang, T. (eds.), \emph{Proceedings of the 38th
  International Conference on Machine Learning}, volume 139 of
  \emph{Proceedings of Machine Learning Research}, pp.\  1964--1974. PMLR,
  18--24 Jul 2021.
\newblock URL \url{https://proceedings.mlr.press/v139/choquette-choo21a.html}.

\bibitem[Christiano et~al.(2017)Christiano, Leike, Brown, Martic, Legg, and
  Amodei]{NIPS2017_d5e2c0ad}
Christiano, P.~F., Leike, J., Brown, T., Martic, M., Legg, S., and Amodei, D.
\newblock Deep reinforcement learning from human preferences.
\newblock In Guyon, I., Luxburg, U.~V., Bengio, S., Wallach, H., Fergus, R.,
  Vishwanathan, S., and Garnett, R. (eds.), \emph{Advances in Neural
  Information Processing Systems}, volume~30. Curran Associates, Inc., 2017.
\newblock URL
  \url{https://proceedings.neurips.cc/paper_files/paper/2017/file/d5e2c0adad503c91f91df240d0cd4e49-Paper.pdf}.

\bibitem[Chung et~al.(2022)Chung, Hou, Longpre, Zoph, Tay, Fedus, Li, Wang,
  Dehghani, Brahma, Webson, Gu, Dai, Suzgun, Chen, Chowdhery, Castro-Ros,
  Pellat, Robinson, Valter, Narang, Mishra, Yu, Zhao, Huang, Dai, Yu, Petrov,
  Chi, Dean, Devlin, Roberts, Zhou, Le, and
  Wei]{chung2022scalinginstructionfinetunedlanguagemodels}
Chung, H.~W., Hou, L., Longpre, S., Zoph, B., Tay, Y., Fedus, W., Li, Y., Wang,
  X., Dehghani, M., Brahma, S., Webson, A., Gu, S.~S., Dai, Z., Suzgun, M.,
  Chen, X., Chowdhery, A., Castro-Ros, A., Pellat, M., Robinson, K., Valter,
  D., Narang, S., Mishra, G., Yu, A., Zhao, V., Huang, Y., Dai, A., Yu, H.,
  Petrov, S., Chi, E.~H., Dean, J., Devlin, J., Roberts, A., Zhou, D., Le,
  Q.~V., and Wei, J.
\newblock Scaling instruction-finetuned language models, 2022.
\newblock URL \url{https://arxiv.org/abs/2210.11416}.

\bibitem[Cui et~al.(2025)Cui, Wei, Swayamdipta, and Jia]{cui-etal-2025-robust}
Cui, X., Wei, J., Swayamdipta, S., and Jia, R.
\newblock Robust data watermarking in language models by injecting fictitious
  knowledge.
\newblock In Che, W., Nabende, J., Shutova, E., and Pilehvar, M.~T. (eds.),
  \emph{Findings of the Association for Computational Linguistics: ACL 2025},
  pp.\  14292--14306, Vienna, Austria, July 2025. Association for Computational
  Linguistics.
\newblock ISBN 979-8-89176-256-5.
\newblock \doi{10.18653/v1/2025.findings-acl.736}.
\newblock URL \url{https://aclanthology.org/2025.findings-acl.736/}.

\bibitem[Dao(2023)]{dao2023flashattention2fasterattentionbetter}
Dao, T.
\newblock Flashattention-2: Faster attention with better parallelism and work
  partitioning, 2023.
\newblock URL \url{https://arxiv.org/abs/2307.08691}.

\bibitem[Dettmers et~al.(2022)Dettmers, Lewis, Shleifer, and
  Zettlemoyer]{dettmers20228bitoptimizersblockwisequantization}
Dettmers, T., Lewis, M., Shleifer, S., and Zettlemoyer, L.
\newblock 8-bit optimizers via block-wise quantization, 2022.
\newblock URL \url{https://arxiv.org/abs/2110.02861}.

\bibitem[Du et~al.(2025)Du, Zhou, Chen, Zhang, Su, Cheng, Chen, and
  Zhang]{du2025sok}
Du, L., Zhou, X., Chen, M., Zhang, C., Su, Z., Cheng, P., Chen, J., and Zhang,
  Z.
\newblock { SoK: Dataset Copyright Auditing in Machine Learning Systems }.
\newblock In \emph{2025 IEEE Symposium on Security and Privacy (SP)}, pp.\
  1--19, Los Alamitos, CA, USA, May 2025. IEEE Computer Society.
\newblock \doi{10.1109/SP61157.2025.00025}.
\newblock URL
  \url{https://doi.ieeecomputersociety.org/10.1109/SP61157.2025.00025}.

\bibitem[Duan et~al.(2024)Duan, Suri, Mireshghallah, Min, Shi, Zettlemoyer,
  Tsvetkov, Choi, Evans, and Hajishirzi]{duan2024membership}
Duan, M., Suri, A., Mireshghallah, N., Min, S., Shi, W., Zettlemoyer, L.,
  Tsvetkov, Y., Choi, Y., Evans, D., and Hajishirzi, H.
\newblock Do membership inference attacks work on large language models?
\newblock In \emph{First Conference on Language Modeling}, 2024.
\newblock URL \url{https://openreview.net/forum?id=av0D19pSkU}.

\bibitem[Duarte et~al.(2024)Duarte, Zhao, Oliveira, and
  Li]{10.5555/3692070.3692545}
Duarte, A.~V., Zhao, X., Oliveira, A.~L., and Li, L.
\newblock {DE-COP}: detecting copyrighted content in language models training
  data.
\newblock In \emph{Proceedings of the 41st International Conference on Machine
  Learning}, ICML'24. JMLR.org, 2024.

\bibitem[Gao et~al.(2020)Gao, Biderman, Black, Golding, Hoppe, Foster, Phang,
  He, Thite, Nabeshima, Presser, and Leahy]{gao2020pile800gbdatasetdiverse}
Gao, L., Biderman, S., Black, S., Golding, L., Hoppe, T., Foster, C., Phang,
  J., He, H., Thite, A., Nabeshima, N., Presser, S., and Leahy, C.
\newblock The pile: An 800{GB} dataset of diverse text for language modeling,
  2020.
\newblock URL \url{https://arxiv.org/abs/2101.00027}.

\bibitem[Geng \& Liu(2023)Geng and Liu]{openlm2023openllama}
Geng, X. and Liu, H.
\newblock {OpenLLaMA}: An open reproduction of {LLaMA}, May 2023.
\newblock URL \url{https://github.com/openlm-research/open_llama}.

\bibitem[Guo et~al.(2025)Guo, Yang, Zhang, et~al.]{Guo_2025}
Guo, D., Yang, D., Zhang, H., et~al.
\newblock {DeepSeek-R1} incentivizes reasoning in {LLMs} through reinforcement
  learning.
\newblock \emph{Nature}, 645\penalty0 (8081):\penalty0 633–638, September
  2025.
\newblock ISSN 1476-4687.
\newblock \doi{10.1038/s41586-025-09422-z}.
\newblock URL \url{http://dx.doi.org/10.1038/s41586-025-09422-z}.

\bibitem[He et~al.(2022)He, Xu, Zeng, Lyu, Wu, Li, and
  Jia]{10.5555/3600270.3600662}
He, X., Xu, Q., Zeng, Y., Lyu, L., Wu, F., Li, J., and Jia, R.
\newblock {CATER}: Intellectual property protection on text generation {API}s
  via conditional watermarks.
\newblock In Oh, A.~H., Agarwal, A., Belgrave, D., and Cho, K. (eds.),
  \emph{Advances in Neural Information Processing Systems}, 2022.
\newblock URL \url{https://openreview.net/forum?id=L7P3IvsoUXY}.

\bibitem[Hu et~al.(2022)Hu, yelong shen, Wallis, Allen-Zhu, Li, Wang, Wang, and
  Chen]{hu2022lora}
Hu, E.~J., yelong shen, Wallis, P., Allen-Zhu, Z., Li, Y., Wang, S., Wang, L.,
  and Chen, W.
\newblock Lo{RA}: Low-rank adaptation of large language models.
\newblock In \emph{International Conference on Learning Representations}, 2022.
\newblock URL \url{https://openreview.net/forum?id=nZeVKeeFYf9}.

\bibitem[Jiang et~al.(2023)Jiang, Sablayrolles, Mensch, Bamford, Chaplot,
  de~las Casas, Bressand, Lengyel, Lample, Saulnier, Lavaud, Lachaux, Stock,
  Scao, Lavril, Wang, Lacroix, and Sayed]{jiang2023mistral7b}
Jiang, A.~Q., Sablayrolles, A., Mensch, A., Bamford, C., Chaplot, D.~S., de~las
  Casas, D., Bressand, F., Lengyel, G., Lample, G., Saulnier, L., Lavaud,
  L.~R., Lachaux, M.-A., Stock, P., Scao, T.~L., Lavril, T., Wang, T., Lacroix,
  T., and Sayed, W.~E.
\newblock Mistral 7b, 2023.
\newblock URL \url{https://arxiv.org/abs/2310.06825}.

\bibitem[Krishna et~al.(2023)Krishna, Song, Karpinska, Wieting, and
  Iyyer]{krishna2023paraphrasing}
Krishna, K., Song, Y., Karpinska, M., Wieting, J.~F., and Iyyer, M.
\newblock Paraphrasing evades detectors of {AI}-generated text, but retrieval
  is an effective defense.
\newblock In \emph{Thirty-seventh Conference on Neural Information Processing
  Systems}, 2023.
\newblock URL \url{https://openreview.net/forum?id=WbFhFvjjKj}.

\bibitem[Lau et~al.(2024)Lau, Niu, Dao, Chen, Foo, and Low]{lau2024waterfall}
Lau, G. K.~R., Niu, X., Dao, H., Chen, J., Foo, C.-S., and Low, B. K.~H.
\newblock Waterfall: Scalable framework for robust text watermarking and
  provenance for {LLM}s.
\newblock In Al-Onaizan, Y., Bansal, M., and Chen, Y.-N. (eds.),
  \emph{Proceedings of the 2024 Conference on Empirical Methods in Natural
  Language Processing}, pp.\  20432--20466, Miami, Florida, USA, November 2024.
  Association for Computational Linguistics.
\newblock \doi{10.18653/v1/2024.emnlp-main.1138}.
\newblock URL \url{https://aclanthology.org/2024.emnlp-main.1138/}.

\bibitem[Li et~al.(2022)Li, Bai, Jiang, Yang, Xia, and Li]{li2022untargeted}
Li, Y., Bai, Y., Jiang, Y., Yang, Y., Xia, S.-T., and Li, B.
\newblock Untargeted backdoor watermark: Towards harmless and stealthy dataset
  copyright protection.
\newblock In Oh, A.~H., Agarwal, A., Belgrave, D., and Cho, K. (eds.),
  \emph{Advances in Neural Information Processing Systems}, 2022.
\newblock URL \url{https://openreview.net/forum?id=kcQiIrvA_nz}.

\bibitem[Liu et~al.(2020)Liu, Ott, Goyal, Du, Joshi, Chen, Levy, Lewis,
  Zettlemoyer, and Stoyanov]{liu2020roberta}
Liu, Y., Ott, M., Goyal, N., Du, J., Joshi, M., Chen, D., Levy, O., Lewis, M.,
  Zettlemoyer, L., and Stoyanov, V.
\newblock {RoBERTa}: A robustly optimized {BERT} pretraining approach, 2020.
\newblock URL \url{https://openreview.net/forum?id=SyxS0T4tvS}.

\bibitem[Liu et~al.(2024)Liu, Hu, Chen, Zhang, and
  Sun]{liu2024watermarkingtextdatalarge}
Liu, Y., Hu, H., Chen, X., Zhang, X., and Sun, L.
\newblock Watermarking text data on large language models for dataset
  copyright, 2024.
\newblock URL \url{https://arxiv.org/abs/2305.13257}.

\bibitem[Loshchilov \& Hutter(2019)Loshchilov and
  Hutter]{loshchilov2018decoupled}
Loshchilov, I. and Hutter, F.
\newblock Decoupled weight decay regularization.
\newblock In \emph{International Conference on Learning Representations}, 2019.
\newblock URL \url{https://openreview.net/forum?id=Bkg6RiCqY7}.

\bibitem[Lu et~al.(2025)Lu, Wang, Zhao, Dai, Foo, Ng, and
  Low]{lu-etal-2025-wasa}
Lu, X., Wang, J., Zhao, Z., Dai, Z., Foo, C.-S., Ng, S.-K., and Low, B. K.~H.
\newblock {WASA}: {WA}termark-based source attribution for large language
  model-generated data.
\newblock In Che, W., Nabende, J., Shutova, E., and Pilehvar, M.~T. (eds.),
  \emph{Findings of the Association for Computational Linguistics: ACL 2025},
  pp.\  23791--23824, Vienna, Austria, July 2025. Association for Computational
  Linguistics.
\newblock ISBN 979-8-89176-256-5.
\newblock \doi{10.18653/v1/2025.findings-acl.1219}.
\newblock URL \url{https://aclanthology.org/2025.findings-acl.1219/}.

\bibitem[Maini et~al.(2024)Maini, Jia, Papernot, and Dziedzic]{maini2024llm}
Maini, P., Jia, H., Papernot, N., and Dziedzic, A.
\newblock {LLM} dataset inference: Did you train on my dataset?
\newblock In \emph{The Thirty-eighth Annual Conference on Neural Information
  Processing Systems}, 2024.
\newblock URL \url{https://openreview.net/forum?id=Fr9d1UMc37}.

\bibitem[Meeus et~al.(2024)Meeus, Shilov, Faysse, and
  De~Montjoye]{10.5555/3692070.3693506}
Meeus, M., Shilov, I., Faysse, M., and De~Montjoye, Y.-A.
\newblock Copyright traps for large language models.
\newblock In \emph{Proceedings of the 41st International Conference on Machine
  Learning}, ICML'24. JMLR.org, 2024.

\bibitem[Meeus et~al.(2025)Meeus, Shilov, Jain, Faysse, Rei, and
  de~Montjoye]{DBLP:conf/satml/MeeusSJFRM25}
Meeus, M., Shilov, I., Jain, S., Faysse, M., Rei, M., and de~Montjoye, Y.
\newblock {SoK}: Membership inference attacks on {LLM}s are rushing nowhere
  (and how to fix it).
\newblock In \emph{{IEEE} Conference on Secure and Trustworthy Machine
  Learning, SaTML 2025, Copenhagen, Denmark, April 9-11, 2025}, pp.\  385--401.
  {IEEE}, 2025.
\newblock \doi{10.1109/SATML64287.2025.00028}.
\newblock URL \url{https://doi.org/10.1109/SaTML64287.2025.00028}.

\bibitem[OpenAI(2024{\natexlab{a}})]{openai2024gpt4ocard}
OpenAI.
\newblock {GPT}-4o system card, 2024{\natexlab{a}}.
\newblock URL \url{https://arxiv.org/abs/2410.21276}.

\bibitem[OpenAI(2024{\natexlab{b}})]{openai2024gpt4technicalreport}
OpenAI.
\newblock {GPT}-4 technical report, 2024{\natexlab{b}}.
\newblock URL \url{https://arxiv.org/abs/2303.08774}.

\bibitem[Ouyang et~al.(2022)Ouyang, Wu, Jiang, Almeida, Wainwright, Mishkin,
  Zhang, Agarwal, Slama, Ray, Schulman, Hilton, Kelton, Miller, Simens, Askell,
  Welinder, Christiano, Leike, and Lowe]{NEURIPS2022_b1efde53}
Ouyang, L., Wu, J., Jiang, X., Almeida, D., Wainwright, C., Mishkin, P., Zhang,
  C., Agarwal, S., Slama, K., Ray, A., Schulman, J., Hilton, J., Kelton, F.,
  Miller, L., Simens, M., Askell, A., Welinder, P., Christiano, P.~F., Leike,
  J., and Lowe, R.
\newblock Training language models to follow instructions with human feedback.
\newblock In Koyejo, S., Mohamed, S., Agarwal, A., Belgrave, D., Cho, K., and
  Oh, A. (eds.), \emph{Advances in Neural Information Processing Systems},
  volume~35, pp.\  27730--27744. Curran Associates, Inc., 2022.
\newblock URL
  \url{https://proceedings.neurips.cc/paper_files/paper/2022/file/b1efde53be364a73914f58805a001731-Paper-Conference.pdf}.

\bibitem[Penedo et~al.(2024)Penedo, Kydl{\'\i}{\v{c}}ek, allal, Lozhkov,
  Mitchell, Raffel, Werra, and Wolf]{10.5555/3737916.3738886}
Penedo, G., Kydl{\'\i}{\v{c}}ek, H., allal, L.~B., Lozhkov, A., Mitchell, M.,
  Raffel, C., Werra, L.~V., and Wolf, T.
\newblock The {FineWeb} datasets: Decanting the web for the finest text data at
  scale.
\newblock In \emph{The Thirty-eight Conference on Neural Information Processing
  Systems Datasets and Benchmarks Track}, 2024.
\newblock URL \url{https://openreview.net/forum?id=n6SCkn2QaG}.

\bibitem[Piet et~al.(2025)Piet, Sitawarin, Fang, Mu, and Wagner]{10992530}
Piet, J., Sitawarin, C., Fang, V., Mu, N., and Wagner, D.
\newblock Markmywords: Analyzing and evaluating language model watermarks.
\newblock In \emph{2025 IEEE Conference on Secure and Trustworthy Machine
  Learning (SaTML)}, pp.\  68--91, 2025.
\newblock \doi{10.1109/SaTML64287.2025.00012}.

\bibitem[Puerto et~al.(2025)Puerto, Gubri, Yun, and
  Oh]{puerto-etal-2025-scaling}
Puerto, H., Gubri, M., Yun, S., and Oh, S.~J.
\newblock Scaling up membership inference: When and how attacks succeed on
  large language models.
\newblock In Chiruzzo, L., Ritter, A., and Wang, L. (eds.), \emph{Findings of
  the Association for Computational Linguistics: NAACL 2025}, pp.\  4165--4182,
  Albuquerque, New Mexico, April 2025. Association for Computational
  Linguistics.
\newblock ISBN 979-8-89176-195-7.
\newblock \doi{10.18653/v1/2025.findings-naacl.234}.
\newblock URL \url{https://aclanthology.org/2025.findings-naacl.234/}.

\bibitem[Raffel et~al.(2020)Raffel, Shazeer, Roberts, Lee, Narang, Matena,
  Zhou, Li, and Liu]{10.5555/3455716.3455856}
Raffel, C., Shazeer, N., Roberts, A., Lee, K., Narang, S., Matena, M., Zhou,
  Y., Li, W., and Liu, P.~J.
\newblock Exploring the limits of transfer learning with a unified text-to-text
  transformer.
\newblock \emph{Journal of Machine Learning Research}, 21\penalty0
  (140):\penalty0 1--67, 2020.
\newblock URL \url{http://jmlr.org/papers/v21/20-074.html}.

\bibitem[Rastogi et~al.(2025)Rastogi, Maini, and Pruthi]{rastogistamp}
Rastogi, S., Maini, P., and Pruthi, D.
\newblock {STAMP} your content: Proving dataset membership via watermarked
  rephrasings.
\newblock In \emph{The 1st Workshop on GenAI Watermarking}, 2025.
\newblock URL \url{https://openreview.net/forum?id=zVbFfPuqPi}.

\bibitem[Sander et~al.(2024)Sander, Fernandez, Durmus, Douze, and
  Furon]{NEURIPS2024_2567c95f}
Sander, T., Fernandez, P., Durmus, A., Douze, M., and Furon, T.
\newblock Watermarking makes language models radioactive.
\newblock In Globerson, A., Mackey, L., Belgrave, D., Fan, A., Paquet, U.,
  Tomczak, J., and Zhang, C. (eds.), \emph{Advances in Neural Information
  Processing Systems}, volume~37, pp.\  21079--21113. Curran Associates, Inc.,
  2024.
\newblock URL
  \url{https://proceedings.neurips.cc/paper_files/paper/2024/file/2567c95fd41459a98a73ba893775d22a-Paper-Conference.pdf}.

\bibitem[Shokri et~al.(2017)Shokri, Stronati, Song, and Shmatikov]{Shokri2017}
Shokri, R., Stronati, M., Song, C., and Shmatikov, V.
\newblock Membership inference attacks against machine learning models.
\newblock In \emph{2017 IEEE Symposium on Security and Privacy (SP)}, pp.\
  3--18, Los Alamitos, CA, USA, May 2017. IEEE Computer Society.
\newblock \doi{10.1109/SP.2017.41}.
\newblock URL \url{https://doi.ieeecomputersociety.org/10.1109/SP.2017.41}.

\bibitem[Soldaini et~al.(2024)Soldaini, Kinney, Bhagia, Schwenk, Atkinson,
  Authur, Bogin, Chandu, Dumas, Elazar, Hofmann, Jha, Kumar, Lucy, Lyu,
  Lambert, Magnusson, Morrison, Muennighoff, Naik, Nam, Peters, Ravichander,
  Richardson, Shen, Strubell, Subramani, Tafjord, Walsh, Zettlemoyer, Smith,
  Hajishirzi, Beltagy, Groeneveld, Dodge, and Lo]{soldaini-etal-2024-dolma}
Soldaini, L., Kinney, R., Bhagia, A., Schwenk, D., Atkinson, D., Authur, R.,
  Bogin, B., Chandu, K., Dumas, J., Elazar, Y., Hofmann, V., Jha, A., Kumar,
  S., Lucy, L., Lyu, X., Lambert, N., Magnusson, I., Morrison, J., Muennighoff,
  N., Naik, A., Nam, C., Peters, M., Ravichander, A., Richardson, K., Shen, Z.,
  Strubell, E., Subramani, N., Tafjord, O., Walsh, E., Zettlemoyer, L., Smith,
  N., Hajishirzi, H., Beltagy, I., Groeneveld, D., Dodge, J., and Lo, K.
\newblock Dolma: an open corpus of three trillion tokens for language model
  pretraining research.
\newblock In Ku, L.-W., Martins, A., and Srikumar, V. (eds.), \emph{Proceedings
  of the 62nd Annual Meeting of the Association for Computational Linguistics
  (Volume 1: Long Papers)}, pp.\  15725--15788, Bangkok, Thailand, August 2024.
  Association for Computational Linguistics.
\newblock \doi{10.18653/v1/2024.acl-long.840}.
\newblock URL \url{https://aclanthology.org/2024.acl-long.840/}.

\bibitem[Tang et~al.(2023)Tang, Feng, Liu, Yang, and
  Hu]{10.1145/3606274.3606279}
Tang, R., Feng, Q., Liu, N., Yang, F., and Hu, X.
\newblock Did you train on my dataset? towards public dataset protection with
  cleanlabel backdoor watermarking.
\newblock \emph{SIGKDD Explor. Newsl.}, 25\penalty0 (1):\penalty0 43–53, July
  2023.
\newblock ISSN 1931-0145.
\newblock \doi{10.1145/3606274.3606279}.
\newblock URL \url{https://doi.org/10.1145/3606274.3606279}.

\bibitem[Weber et~al.(2024)Weber, Fu, Anthony, Oren, Adams, Alexandrov, Lyu,
  Nguyen, Yao, Adams, Athiwaratkun, Chalamala, Chen, Ryabinin, Dao, Liang,
  R\'{e}, Rish, and Zhang]{10.5555/3737916.3741613}
Weber, M., Fu, D.~Y., Anthony, Q., Oren, Y., Adams, S., Alexandrov, A., Lyu,
  X., Nguyen, H., Yao, X., Adams, V., Athiwaratkun, B., Chalamala, R., Chen,
  K., Ryabinin, M., Dao, T., Liang, P., R\'{e}, C., Rish, I., and Zhang, C.
\newblock {RedPajama}: an open dataset for training large language models.
\newblock In Globerson, A., Mackey, L., Belgrave, D., Fan, A., Paquet, U.,
  Tomczak, J., and Zhang, C. (eds.), \emph{Advances in Neural Information
  Processing Systems}, volume~37, pp.\  116462--116492. Curran Associates,
  Inc., 2024.
\newblock \doi{10.52202/079017-3697}.
\newblock URL
  \url{https://proceedings.neurips.cc/paper_files/paper/2024/file/d34497330b1fd6530f7afd86d0df9f76-Paper-Datasets_and_Benchmarks_Track.pdf}.

\bibitem[Wei et~al.(2024)Wei, Wang, and Jia]{wei-etal-2024-proving}
Wei, J., Wang, R., and Jia, R.
\newblock Proving membership in {LLM} pretraining data via data watermarks.
\newblock In Ku, L.-W., Martins, A., and Srikumar, V. (eds.), \emph{Findings of
  the Association for Computational Linguistics: ACL 2024}, pp.\  13306--13320,
  Bangkok, Thailand, August 2024. Association for Computational Linguistics.
\newblock \doi{10.18653/v1/2024.findings-acl.788}.
\newblock URL \url{https://aclanthology.org/2024.findings-acl.788/}.

\bibitem[Wenger et~al.(2024)Wenger, Li, Zhao, and Shmatikov]{wenger2022data}
Wenger, E., Li, X., Zhao, B.~Y., and Shmatikov, V.
\newblock Data isotopes for data provenance in dnns.
\newblock \emph{Proc. Priv. Enhancing Technol.}, 2024\penalty0 (1):\penalty0
  413--429, January 2024.
\newblock URL \url{https://doi.org/10.56553/popets-2024-0024}.

\bibitem[Wenzek et~al.(2020)Wenzek, Lachaux, Conneau, Chaudhary, Guzm{\'a}n,
  Joulin, and Grave]{wenzek-etal-2020-ccnet}
Wenzek, G., Lachaux, M.-A., Conneau, A., Chaudhary, V., Guzm{\'a}n, F., Joulin,
  A., and Grave, E.
\newblock {CCN}et: Extracting high quality monolingual datasets from web crawl
  data.
\newblock In Calzolari, N., B{\'e}chet, F., Blache, P., Choukri, K., Cieri, C.,
  Declerck, T., Goggi, S., Isahara, H., Maegaard, B., Mariani, J., Mazo, H.,
  Moreno, A., Odijk, J., and Piperidis, S. (eds.), \emph{Proceedings of the
  Twelfth Language Resources and Evaluation Conference}, pp.\  4003--4012,
  Marseille, France, May 2020. European Language Resources Association.
\newblock ISBN 979-10-95546-34-4.
\newblock URL \url{https://aclanthology.org/2020.lrec-1.494/}.

\bibitem[Zhang et~al.(2025)Zhang, Das, Kamath, and Tramer]{10992516}
Zhang, J., Das, D., Kamath, G., and Tramer, F.
\newblock { Position: Membership Inference Attacks Cannot Prove That a Model
  was Trained on Your Data }.
\newblock In \emph{2025 IEEE Conference on Secure and Trustworthy Machine
  Learning (SaTML)}, pp.\  333--345, Los Alamitos, CA, USA, April 2025. IEEE
  Computer Society.
\newblock \doi{10.1109/SaTML64287.2025.00025}.
\newblock URL
  \url{https://doi.ieeecomputersociety.org/10.1109/SaTML64287.2025.00025}.

\bibitem[Zhang et~al.(2022)Zhang, Roller, Goyal, Artetxe, Chen, Chen, Dewan,
  Diab, Li, Lin, Mihaylov, Ott, Shleifer, Shuster, Simig, Koura, Sridhar, Wang,
  and Zettlemoyer]{zhang2022optopenpretrainedtransformer}
Zhang, S., Roller, S., Goyal, N., Artetxe, M., Chen, M., Chen, S., Dewan, C.,
  Diab, M., Li, X., Lin, X.~V., Mihaylov, T., Ott, M., Shleifer, S., Shuster,
  K., Simig, D., Koura, P.~S., Sridhar, A., Wang, T., and Zettlemoyer, L.
\newblock {OPT}: Open pre-trained transformer language models, 2022.
\newblock URL \url{https://arxiv.org/abs/2205.01068}.

\bibitem[Zhao et~al.(2022)Zhao, Li, and Wang]{DBLP:conf/emnlp/ZhaoLW22}
Zhao, X., Li, L., and Wang, Y.
\newblock Distillation-resistant watermarking for model protection in {NLP}.
\newblock In Goldberg, Y., Kozareva, Z., and Zhang, Y. (eds.), \emph{Findings
  of the Association for Computational Linguistics: {EMNLP} 2022, Abu Dhabi,
  United Arab Emirates, December 7-11, 2022}, pp.\  5044--5055. Association for
  Computational Linguistics, 2022.
\newblock \doi{10.18653/V1/2022.FINDINGS-EMNLP.370}.
\newblock URL \url{https://doi.org/10.18653/v1/2022.findings-emnlp.370}.

\bibitem[Zhao et~al.(2023)Zhao, Wang, and Li]{10.5555/3618408.3620182}
Zhao, X., Wang, Y.-X., and Li, L.
\newblock Protecting language generation models via invisible watermarking.
\newblock In \emph{Proceedings of the 40th International Conference on Machine
  Learning}, ICML'23. JMLR.org, 2023.

\bibitem[Zhao et~al.(2025{\natexlab{a}})Zhao, Gunn, Christ, Fairoze, Fabrega,
  Carlini, Garg, Hong, Nasr, Tramer, Jha, Li, Wang, and Song]{sokgenai}
Zhao, X., Gunn, S., Christ, M., Fairoze, J., Fabrega, A., Carlini, N., Garg,
  S., Hong, S., Nasr, M., Tramer, F., Jha, S., Li, L., Wang, Y.-X., and Song,
  D.
\newblock { SoK: Watermarking for AI-Generated Content }.
\newblock In \emph{2025 IEEE Symposium on Security and Privacy (SP)}, pp.\
  2621--2639, Los Alamitos, CA, USA, May 2025{\natexlab{a}}. IEEE Computer
  Society.
\newblock \doi{10.1109/SP61157.2025.00178}.
\newblock URL
  \url{https://doi.ieeecomputersociety.org/10.1109/SP61157.2025.00178}.

\bibitem[Zhao et~al.(2025{\natexlab{b}})Zhao, Liu, Jha, McDaniel, Li, and
  Xiao]{zhao2025can}
Zhao, Z., Liu, X., Jha, S., McDaniel, P., Li, B., and Xiao, C.
\newblock Can watermarks be used to detect {LLM} {IP} infringement for free?
\newblock In Yue, Y., Garg, A., Peng, N., Sha, F., and Yu, R. (eds.),
  \emph{International Conference on Learning Representations}, volume 2025,
  pp.\  57060--57076, 2025{\natexlab{b}}.
\newblock URL
  \url{https://proceedings.iclr.cc/paper_files/paper/2025/file/8fba406323cb3930aeaccc9aa64c83a8-Paper-Conference.pdf}.

\end{thebibliography}

\appendix
\onecolumn
\section{Detailed algorithms and analysis}
\label{algs}

This section details algorithms applied in this proposal. The notations used in the paper are recalled in table~\ref{tab:notations}.

\begin{table}[htbp]

\caption{Notations}
\begin{center}
    \begin{small}
\label{tab:notations}
\begin{tabular}{p{0.2\textwidth}|p{0.6\textwidth}}
  \toprule
  $\col = \{\doc\}$ & A sensitive collection of documents    \\
 \midrule
  $\doc = (\word_i)_{ i = 1..\Delta}$ & A document of $\Delta$ words   \\
  \midrule
  $\mathcal{L}$ & A suspicious model \\
  \midrule
  $\mathcal{P} = (verif,mark)$ & A dataset provenance auditing procedure \\
    \midrule
  $\mathcal{A}$ & The alphabet of invisible characters   \\
  \midrule
 $\syllable \in \mathcal{A}^\sSy$ & A syllable of \sSy invisible character  \\
 \midrule
  $\ws$ & The domain of watermarks  \\
 \midrule
 $\ws_\lo \subset \ws$ & Set of $\lo$ candidate watermarks \\ & provided upon request\\ 
 \midrule
 $\w = (\syllable_i)_{ i=1..\sW}$  & A watermark of $\sW$ syllables \\ 
 \midrule
$\w c= (\syllable_i)_{ \syllable_i \in w, i \leq\sCu}$  & The ``cue'' of $\w$ made of its first $\sCu$ syllables \\
 \midrule
$(\syllable_i)_{ \syllable_i \in w, \sCu -t < i \leq\sCu}$  & The tail of the ``cue'', composed of its $\cinr$ last syllables \\
\midrule
$\w r= (\syllable_i)_{\syllable_i \in w, i > \sCu}$  & the ``reply'' of $\w$ made  of its last $\sW - \sCu$ syllables \\
\midrule
\spacing & The ``step'' in the number of words used in the marking scheme \\ 
\midrule
  $\doc = {(\chunk)}_{i=1..\lceil\Delta/\delta\rceil}$ & $\doc$ is a sequence of contiguous  chunks of $\delta$ words \\
\midrule
$\{\doc^{w c}_{2i+1}\}$ & Odd chunks of $\doc^\w$ watermarked with syllables of cue not in its tail  \\
\midrule
$\{\doc^{w r}_{2i}\}$ & Even chunks of $\doc^\w$ watermarked with  syllables of reply and of the tail of cue   \\
\midrule
  $\doc^{\w} = mark(\doc, \w)$ & $\doc$ chunks marked by scheme  $mark$ using $\w$  \\
  \midrule
  $ID^w_i$ & Challenge constructed from $D^{cw}_{2i-1}$ \\ \midrule
  $\lambda$ & Number of repetitions of each $ID^w_i$ during $verif$ \\   \bottomrule
\end{tabular}
    \end{small}
  \end{center}
  \vskip -0.1in
\end{table}

\subsection{Number of admissible watermarks}

We assume $\sCu \geq \sW - \sCu$, i.e., that the \cue is longer than the \sign. 
A given sequence of $\sSy\sCu$ characters appears in $(\sSy(2\sCu - \sW) +1) \times |\mathcal{A}|^{\sSy(2\sCu - \sW)}$ sequence of $\sSy(\sW - \sCu)$ characters at most. 

For instance, with $\sCu$ = 2, $\sW - \sCu$ = 1, and  $\sSy = 2$, a sequence of 2 characters, e.g. ``aa'', may appear in $(\sSy(2\sCu - \sW) +1)$ = 3 different positions in a sequence of 4 characters, e.g. ``aa**'', ``*aa*'', or ``**aa''. There remains $\sSy(2\sCu - \sW)$ = 2 other characters that may take any value in $|\mathcal{A}|$. 

Hence, considering a particular \sign forbids the usage of at most $(\sSy(2\sCu - \sW) +1) \times |\mathcal{A}|^{\sSy(2\sCu - \sW)}$ distinct \cue. This is an upper bound as multiple occurrences of a \sign in a single \cue are counted multiple time, e.g. ``aaaa'' is counted twice.

Hence, due to constraint (2), each used \sign ``forbids'' at most $(\sSy(2\sCu - \sW) +1) \times |\mathcal{A}|^{\sSy(2\sCu - \sW)}$ \cue. 

Because of condition (1), the number of distinct watermarks equals the minimum of the number of distinct \cue and \sign. Hence, we have,

\[
 |\ws| \geq max_{x \in [1,|\mathcal{A}|^{\sSy(\sW-\sCu)}]}  min(  x,  |\mathcal{A}|^{\sSy\sCu}-x\times(\sSy(2\sCu - \sW) +1)  |\mathcal{A}|^{\sSy(2\sCu - \sW)})
\]

Note that, if $\sCu = \sW - \sCu$, i.e., the \cue and \sign have the same length, (2) reduces to an inequality and the optimum is reached when the space is equally partitioned between \sign and \cue. As expected, by substitution in the above equation, we get:

\[
 |\ws| \geq max_{x \in [1,|\mathcal{A}|^{\sSy\sCu}]}  min(  x,  |\mathcal{A}|^{\sSy\sCu}-x\})
 \quad \text{i.e. } |\ws| \geq \frac{|\mathcal{A}|^{\sSy\sCu}}{2} 
\]

\subsection{Embedding}
\label{app:embed}

Algorithm \ref{alg:embedfull} presents the embedding process. It chunks \doc into $\lceil\Delta/\delta\rceil$ sub-documents ${(\chunk)}_{i=1..\lceil\Delta/\delta\rceil}$ containing $\delta$ words (except for the last). Alternatively, a chunk is embedded with either (most of) the \cue, $(\syllable_i)_{s_i \in w, i \leq \sCu}$, or $(\syllable_i)_{s_i \in w, i \geq \sCu-\cinr}$ (i.e. the \cinr\ last syllables of the \cue and the \sign). Sub-documents with odd indexes, or \cue chunks, are denoted $\{\doc^{w c}_{2i+1}\}$. Sub-documents with even indexes or \sign chunks, are denoted $\{\doc^{w r}_{2i}\}$.

The function \textit{chunk\_mark} embeds the relevant syllables within a subset of the document. The first syllable is inserted after the first word, and the remaining ones are inserted every \spacing word. The $\sCu-\cinr$ or $\sW -\sCu+\cinr$ syllables are injected cyclically throughout the list. When the end of this list is reached, a syllable is not necessarily inserted even if the last one was injected exactly \spacing words before. Rather, if the final repetition is incomplete, the remaining syllables are concatenated at the end to ensure full representation.

\begin{algorithm}[!htbp]
\caption{\textit{mark}, embed Watermark $\w$ in $\doc$}
\label{alg:embedfull}
\begin{algorithmic}[1]
 \renewcommand{\algorithmicrequire}{\textbf{Input: $\doc= (\word_i)_{i=1..\lceil\Delta/\delta\rceil}$, $\delta$, \cinr, \spacing, \sCu, $\w$}}
 \renewcommand{\algorithmicensure}{\textbf{Output: $\doc^\w$, $\doc$ watermarked with $\w$}}
 \REQUIRE 
 \ENSURE
 \STATE $D^\w \xleftarrow{}$ empty list
 \STATE $x\leftarrow{}$ 1 \COMMENT{Start of chunk}
 \WHILE{$x + \delta < \Delta$}
   \STATE $y\leftarrow{} x+\delta -1$
  \STATE $\doc^{\w c} \xleftarrow{} \text{chunk\_mark}((\word_i)_{i=x..y},(\syllable_i)_{s_i \in \w, i \leq \sCu - \cinr})$
  \STATE Concatenate $\doc^\w$ and $\doc^{w c}$
   \STATE $x\leftarrow{} x + \lambda $  
\STATE $y\leftarrow{} min(x+\delta -1,\Delta)$
\STATE $\doc^{wr} \xleftarrow{} \text{chunk\_mark}((\word_i)_{i=x..y}, (\syllable_i)_{s_i \in w, i > \sCu - \cinr})$
\STATE Concatenate $\doc^\w$ and $\doc^{wr}$ 
\STATE $x\xleftarrow{} x+ \delta  $
\ENDWHILE
\IF{$x\neq \Delta$} 
\STATE Concatenate $\doc^\w$ and $(\word_i)_{i=x..\Delta}$ 
\ENDIF
\RETURN $\doc^\w$

\hspace{-0,6cm}\textbf{Subroutine} \textit{chunk\_mark}
 \renewcommand{\algorithmicrequire}{\textbf{Input: Words $(\word_i)_{i=1..N}$, syllables $(\syllable_i)_{i=1..M}$}}
 \renewcommand{\algorithmicensure}{\textbf{Output: Sequence $\doc^\w$ of words and syllables}}
 \REQUIRE 
 \ENSURE
 \STATE $\doc^\w \xleftarrow{}$ empty list 
 \STATE Append $\word_1$ to $\doc^\w$ 
 \STATE Append $\syllable_1$ to $\doc^\w$ 
  \STATE $x \xleftarrow{} 2$ \COMMENT{word count}
 \STATE $y \xleftarrow{} 2$ \COMMENT{syllable count}
 \WHILE{$x + \spacing  \leq N$}
 \STATE Concatenate $\doc^\w$ and $(\word_i)_{i=x..x+\spacing-1}$
 \STATE Append $\syllable_y$  to $\doc^\w$
 \STATE $x \xleftarrow{} x + \spacing$
 \STATE  $y \xleftarrow{} (y \bmod M)+1$ \COMMENT{Cycle through syllables}
 \ENDWHILE
 \STATE Concatenate $\doc^\w$ and $(\word_i)_{i =x..N}$ 
 \WHILE{y $\neq$ 1}
 \STATE Append $\syllable_y$  to $\doc^\w$
 \STATE $y \xleftarrow{} (y\bmod M)+1$
 \ENDWHILE
 \RETURN $\doc^\w$
\end{algorithmic}
\end{algorithm}

Following the strategy described in Algorithm~\ref{alg:embedfull}, if $y$ is the number of syllables taken as input and $x$ is the number of words, $1 + \lfloor \frac{x-2}{\spacing} \rfloor$  are embedded first. Then,  any remaining syllables are appended at the end to ensure complete representation. The total number of repetitions in a sub-document is thus:
\[
      \text{sub-repetitions(x,y)} =
 \left\lceil 
 \frac{1 + \left\lfloor \frac{x - 2}{\spacing} \right\rfloor}{y} 
\right\rceil
\]

Recall that odd subdocuments are embedded with the $\sCu-\cinr$ first syllables of the  \cue, the remaining $\cinr$ being embedded in even subdocuments alongside the $\sW - \sCu$ syllables of the \sign.  A document of size $\Delta$ has:
\begin{itemize}
    \item $\lceil\Delta/\delta\rceil$/2 sub-document of size $\delta $ embedded with a signal of $\sCu - \cinr$ syllables, 
    \item $\lceil\Delta/\delta\rceil$/2-1 sub-documents of size $\delta$ embedded with a signal of $\sW - \sCu +\cinr$ syllables, and
    \item a final sub-document of size $\Delta - \delta\cdot(\lceil \Delta/\delta \rceil-1)$ embedded with a signal of $\sW - \sCu +\cinr$ syllables.
\end{itemize} 

Thus, the number of repetitions of $(\syllable_i)_{s_i \in w, i \leq \sCu -\cinr}$  in \doc is:
\[
 \lceil\Delta/\delta\rceil/2 \times
 \left\lceil 
 \frac{1 + \left\lfloor \frac{\delta  - 2}{\spacing} \right\rfloor}{\sCu - \cinr} 
\right\rceil
\]

And the number of repetitions of $(\syllable_i)_{s_i \in w, i > \sCu - \cinr}$ in \doc is:
\[
 \lceil\Delta/\delta\rceil/2 \times
 \left\lceil 
 \frac{1 + \left\lfloor \frac{\delta - 2}{\spacing} \right\rfloor}{\sW - \sCu +\cinr} 
\right\rceil +  \left\lceil 
 \frac{1 + \left\lfloor \frac{\Delta - \delta\cdot(\lceil \Delta/\delta \rceil-1) - 2}{\spacing} \right\rfloor}{\sW - \sCu + \cinr} 
\right\rceil
\]

\subsection{Eliciting and detecting a \sign}

We denote by $\mathcal{L}(I)$ the output generated by a LLM $\mathcal{L}$ when prompted with input $I$. This output is treated as an ordered list of words and invisible characters. Unlike the embedding stage, these extracted invisible characters are not assumed to be in a syllable of size \sSy.  For a watermarked document $\doc^\w$, the verification consists of constructing $\lfloor\frac{\lceil \Delta/\delta \rceil}{2}\rfloor$ inputs $(ID^\w_i)_{i=1..\lfloor\frac{\lceil \Delta/\delta \rceil}{2}\rfloor}$ made of:

\begin{itemize}
    \item $\doc^{\w c}_{2(i-1)+1}$, containing the repeated $\sCu-\cinr$ syllables of the \cue $\w c$, followed by
    \item the first $\cinr\cdot(1+\spacing)$ elements of $\doc^{\w r}_{2i}$ (i.e. the first $\spacing\cdot\cinr$ words of $\doc^{\w r}_{2i}$ embedded with the tail of the \cue $\w c$). This completes the partial \cue of $\doc^{\w c}_{2(i-1)+1}$ and serves as a stimulus for the appearance of the \sign $\w r$.
\end{itemize}

We then query the model $\lambda$ times with each $ID^\w_i$. The chunk-level verification, described in Algorithm~\ref{alg:challenge} outputs one if and only if the \sign $\w r$ is detected in $\mathcal{L}(ID_\w^i)$ in at least one of the $\lambda$ repetitions. Detection is conducted by filtering the model's output, removing all visible characters, and concatenating only the invisible ones in their original order into $o_{\text{invis}}$. If any subsequence of $o_{\text{invis}}$ matches $\w r$, the \sign has been detected.

\begin{algorithm}[H]
\caption{$verif$, searching for \sign in $\mathcal{L}(ID^w_i$))}
\label{alg:challenge}
\begin{algorithmic}[1]
 \renewcommand{\algorithmicrequire}{\textbf{Input: LLM $\mathcal{L}$, a prompt $ID^w_i$}, repetition factor $\lambda$}
 \renewcommand{\algorithmicensure}{\textbf{Output: 1 if the \sign is detected, 0 else}}
 \REQUIRE 
 \ENSURE
 \FOR{$x=1..\lambda$}
 \STATE $o \xleftarrow[]{}$  $\mathcal{L}(ID^w_i)$
  \STATE $o_{\text{invis}} \xleftarrow{}$ empty list
 \FOR{$o_i \in o$}
 \IF{$o_i \cap \mathcal{A} \neq \emptyset$}
 \FORALL{$a \in o_i \cap \mathcal{A}$ in the order they appear in $o_i$}
 \STATE Append $a$ to $o_{\text{invis}}$
 \ENDFOR
 \ENDIF
 \ENDFOR
 \STATE Flatten $\w r$ into a single sequence of invisible characters $\w^{r}_{\text{flat}}$
\IF{$\w^{r}_{\text{flat}}$ is a contiguous subsequence of $o_{\text{invis}}$}
\RETURN 1
\ENDIF
\ENDFOR
\RETURN 0

\end{algorithmic}
\end{algorithm}

\subsection{Ranking test to assess membership}

Algorithm~\ref{alg:test} describes the ranking test procedure to determine membership. 
We do not assume that the collection was either wholly used or wholly unused during fine-tuning. Instead, we aim to determine whether elements of the collection have been used.

\begin{algorithm}[H]
\caption{$decision$, assess membership}
\label{alg:test}
\begin{algorithmic}[1]
 \renewcommand{\algorithmicrequire}{\textbf{Input: $\mathcal{L}, \col, \text{rank } k, \lambda, \w, \ws_\lo$}}
 \renewcommand{\algorithmicensure}{\textbf{Output: $\exists d^\w \in \col^\w$ on which $\mathcal{L}$ was trained}}
 \REQUIRE 
 \ENSURE
 \STATE $s \xleftarrow{} verif(\mathcal{L},\doc^{\w},\lambda)$
 \STATE rank $\xleftarrow{}$ 1 
 \FORALL{$\w_\lo \in \ws_\lo$}
 \STATE $s_\lo \xleftarrow{} verif(\mathcal{L},\doc^{\w_\lo},\lambda)$
 \IF{$s_\lo \geq s$}
 \STATE rank $\xleftarrow{}$ rank + 1 
 \ENDIF
 \ENDFOR
 \RETURN rank $\leq$ k
 \end{algorithmic}
\end{algorithm}

\clearpage
\section{Experiment Details}
\label{sec:xpdetails}
Table~\ref{tab: XP details} presents the experimental setup, including the low-rank adaptation (LoRA; \citealp{hu2022lora}) configuration, training parameters, and generation settings for each experiment. All experiments were conducted on NVIDIA H100 and H200 GPUs under identical experimental settings.

\begin{table}[!htbp]
\centering
\caption{LoRA configuration and training setup.}
\label{tab: XP details}
\begin{tabularx}{\linewidth}{l X}
\hline
\textbf{Parameter} & \textbf{Value} \\
\hline
\multicolumn{2}{l}{\textbf{LoRA configuration}} \\
\hline
Target modules & q\_proj, k\_proj, v\_proj, o\_proj \\
Rank ($r$) & 12 \\
Alpha & 32 \\
Dropout & 0.05 \\
Bias & none \\
Task type & Causal LM \\
\hline
\multicolumn{2}{l}{\textbf{Training setup}} \\
\hline
Base model & \href{https://huggingface.co/mistralai/Mistral-7B-v0.1}{Mistral-7B-v0.1}, \href{https://huggingface.co/meta-LLaMA/LLaMA-2-7b-hf}{LLaMA-2-7B-hf} and  \href{https://huggingface.co/openai/gpt-oss-20b}{GPT-OSS-20B}\\
Dataset split & 90\% train / 10\% eval \\
Sequence length & 4096 tokens \\
Attention & FlashAttention 2~\citep{dao2023flashattention2fasterattentionbetter} for Mistral and LLaMA and eager for GPT-OSS\\
Precision & bfloat16 for Mistral and LLaMA and MXFP4 for GPT-OSS\\
Optimizer & paged\_adamw\_8bit~\citep{loshchilov2018decoupled, dettmers20228bitoptimizersblockwisequantization} \\
Learning rate & $2\times 10^{-4}$ (cosine schedule) \\
Warmup ratio & 0.03 \\
Weight decay & 0.05 \\
Batch size & 2 per device \\
Grad. accumulation & 8 steps (effective batch size 16) \\
Epochs & 3 \\
\hline
\multicolumn{2}{l}{\textbf{Generation configuration}} \\
\hline
Samlpe & True \\
Temperature & 0.7 \\
Top p & 0.9 \\
Top k & 50 \\
Max new tokens & 200 \\
\hline
\end{tabularx}
\end{table}

\clearpage

\section{Watermarked Text Statistics}
\label{sec:stats}

Table~\ref{tab:statsxp1&2} summarizes basic length statistics of the watermarked documents used for finetuning in Sections~\ref{sec: Impact of the size of the watermarked collection} and~\ref{sec: Impact of the size of the dataset}. 
For each dataset and each configuration, we report the number of watermarked documents as well as the minimum, maximum, mean, and standard deviation of document length measured in number of words. 

In the experiments of Section~\ref{sec: Impact of the size of the dataset}, only the size of the overall training set is varied, while the set of watermarked documents is kept fixed. Consequently, all configurations in that setting share identical watermarked texts and therefore identical statistics in Table~\ref{tab:statsxp1&2}.

\begin{table}[H]
\centering
\caption{Statistics of watermarked documents in the \textbf{Poems}, \textbf{Blog}, and \textbf{News} datasets. The last column approximates the number of watermark repetitions per document, assuming one occurence per 32 words.}
\label{tab:statsxp1&2}
\centering
\begin{tabularx}{0.7\linewidth}{l c c c c c c}

\hline
\textbf{Dataset} & \textbf{\#WMed Docs} & \multicolumn{4}{c}{\textbf{Doc. Length}}  & \textbf{$\approx$ WM rep./Doc} \\
& & \textbf{Min} & \textbf{Max} & \textbf{Mean} & \textbf{Std} & \\
\hline
\hline
\multicolumn{7}{l}{\textbf{Poems}} \\
\hline
 & 10  & 214  & 532  & 336.0  & 102.0 & 10 \\
 & 20  & 203  & 592  & 342.9  & 105.4 & 11 \\
 & 30  & 203  & 592  & 331.0  & 104.5 & 10 \\
 & 40  & 202  & 809  & 325.0  & 128.0 & 10 \\
 & 50  & 202  & 914  & 333.9  & 147.0 & 10 \\
 & 60  & 202  & 914  & 334.8  & 140.9 & 10 \\
 & 70  & 202  & 914  & 339.4  & 136.1 & 11 \\
 & 80  & 202  & 914  & 338.9  & 145.7 & 11 \\
 & 90  & 202  & 914  & 333.3  & 141.8 & 10 \\
 & 100 & 202  & 1996 & 349.1  & 222.2 & 11 \\
\hline
\multicolumn{7}{l}{\textbf{Blog}} \\
\hline
 & 10  & 252  & 934  & 417.9  & 181.0 & 13 \\
 & 20  & 222  & 934  & 383.1  & 151.5 & 12 \\
 & 30  & 215  & 1264 & 400.7  & 213.2 & 13 \\
 & 40  & 204  & 1264 & 377.3  & 194.7 & 12 \\
 & 50  & 204  & 1264 & 387.2  & 200.5 & 12 \\
 & 60  & 204  & 1264 & 375.1  & 188.0 & 12 \\
 & 70  & 203  & 1264 & 361.9  & 179.6 & 11 \\
 & 80  & 203  & 1836 & 370.5  & 238.5 & 12 \\
 & 90  & 203  & 1836 & 361.6  & 228.0 & 11 \\
 & 100 & 203  & 1836 & 357.3  & 219.3 & 11 \\
\hline
\multicolumn{7}{l}{\textbf{News}} \\
\hline
 & 1   & 1665 & 1665 & 1665.0  & 0.0 & 52 \\
 & 5   & 1427 & 1743 & 1570.0  & 119.2 & 49 \\
 & 10  & 1369 & 1743 & 1506.4  & 118.2 & 47 \\
 & 20  & 1329 & 1743 & 1508.0  & 134.8 & 47 \\
 & 30  & 1270 & 1743 & 1492.7  & 124.9 & 47 \\
 & 40  & 1240 & 1743 & 1471.7  & 126.3 & 46 \\
 & 50  & 1240 & 1743 & 1475.8  & 125.8 & 46 \\
 & 60  & 1240 & 1743 & 1484.0  & 122.2 & 46 \\
 & 70  & 1240 & 1743 & 1479.6  & 122.0 & 46 \\
 & 80  & 1240 & 1743 & 1484.1  & 123.0 & 46 \\
 & 90  & 1240 & 1743 & 1480.5  & 120.6 & 46 \\
 & 100 & 1240 & 1743 & 1475.3  & 120.7 & 46 \\
\hline
\end{tabularx}
\end{table}

\clearpage

Table~\ref{tab: statistics xp3} presents the statistics of the watermarked texts used in finetuning in Section~\ref{sec: Impact of the number of unique watermarks} as well as in the additional experiment conducted using GPT-OSS. In these experiments, the number of documents associated with each unique watermark was kept fixed, while only the number of unique watermarks ($U$) was varied. GPT-OSS uses a larger number of documents per unique watermark (100) to account for its increased model size.

\begin{table}[!htbp]
\centering
\small
\caption{Statistics of watermarked documents with varying numbers of unique watermarks ($U$). The last column approximates the number of watermarks per document, assuming one watermark per 32 words.}
\label{tab: statistics xp3}
\begin{tabularx}{0.74\linewidth}{l c c c r r r c}
\hline
\textbf{Data} & \textbf{Model} & $U$ & \textbf{Doc/WM}  & \multicolumn{3}{c}{\textbf{Doc. Length}} & \textbf{$\approx$ WM rep./Doc} \\
& & & &  \textbf{Min} & \textbf{Max} & \textbf{Mean} & \\
\hline
\hline
 \textbf{Blog}& LLaMA & 5  & 40  & 200  & 1357 & 352.8 & 11 \\
 & LLaMA & 20 & 40 & 200  & 2743 & 368.3 & 12 \\
 & LLaMA & 30 & 40  & 200  & 2743 & 364.0 & 11 \\ 
 & LLaMA & 40 & 40  & 200  & 2743 & 361.5 & 11 \\
 & LLaMA & 50 & 40  & 200  & 2743 & 359.9 & 11 \\ 
 & Mistral & 5  & 60   & 200  & 1357 & 359.1 & 11 \\
 & Mistral & 20 & 60  & 200  & 2743 & 363.3 & 11 \\
 & Mistral & 30 & 60  & 200  & 2743 & 361.5 & 11 \\
 & Mistral & 40 & 60  & 200  & 2743 & 360.1 & 11 \\
 & Mistral & 50 & 60  & 200  & 2743 & 357.5 & 11 \\
 & GPT-OSS & 5 & 100  & 200 & 1836 & 354.6 & 11 \\
 & GPT-OSS & 20 & 100  & 200 & 2743 & 357.8 & 11 \\
 & GPT-OSS & 30 & 100 & 200 & 2743 & 358.8 & 11 \\
 & GPT-OSS & 40 & 100  & 200 & 2743 & 358.4 & 11 \\
 & GPT-OSS & 50 & 100  & 200 & 2743 & 356.3 & 11 \\
\hline
\hline
 \textbf{Poems}& Mistral/LLaMA & 5  & 40  & 200  & 3172 & 424.4 & 13 \\
 & Mistral/LLaMA & 20 & 40   & 200  & 9722 & 450.1 & 14 \\
 & Mistral/LLaMA & 30 & 40  & 200  & 9722 & 454.9 & 14 \\
 & Mistral/LLaMA & 40 & 40  & 200  & 9722 & 499.1 & 16 \\
 & Mistral/LLaMA & 50 & 40  & 200  & 9722 & 491.6 & 15 \\
 & GPT-OSS & 5 & 100 & 200 & 8391 & 525.5 & 16 \\
 & GPT-OSS & 20 & 100  & 200 & 9722 & 460.3 & 14 \\
 & GPT-OSS & 30 & 100  & 200 & 9722 & 460.3 & 14 \\
 & GPT-OSS & 40 & 100  & 200 & 9722 & 484.9 & 15 \\
 & GPT-OSS & 50 & 100  & 200 & 9722 & 469.1 & 15 \\
\hline
\hline
 \textbf{News}& Mistral/LLaMA & 20 & 40   & 1218 & 1803 & 1486.8 & 46 \\
 & Mistral/LLaMA & 30 & 40  & 1218 & 1803 & 1484.2 & 46 \\
 & Mistral/LLaMA & 40 & 40  & 1218 & 1803 & 1488.5 & 47 \\
 & Mistral/LLaMA & 50 & 40  & 1218 & 1836 & 1500.0 & 47 \\
 & GPT-OSS & 5 & 100 & 1240 & 1754 & 1512.2 & 47 \\
 & GPT-OSS & 20 & 100  & 1218 & 1803 & 1483.6 & 46 \\
 & GPT-OSS & 30 & 100  & 1218 & 1803 & 1479.1 & 46 \\
 & GPT-OSS & 40 & 100 & 1218 & 1833 & 1485.8 & 46 \\
 & GPT-OSS & 50 & 100  & 881 & 1836 & 1497.1 & 47 \\
 \hline
\end{tabularx}
\end{table}

\clearpage

\section{Per-watermark regurgitation rate in multi-watermark settings}
\label{app: Details of unique watermarks experiment}

\begin{figure}[!h]
    \centering

    \begin{subfigure}[t]{0.42\textwidth}
        \centering
        \includegraphics[width=\linewidth]{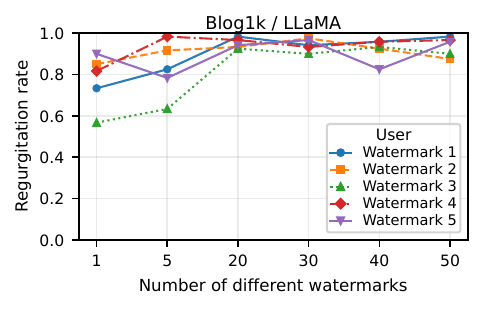}
        \label{fig:big_a}
    \end{subfigure}\hfill
    \begin{subfigure}[t]{0.42\textwidth}
        \centering
        \includegraphics[width=\linewidth]{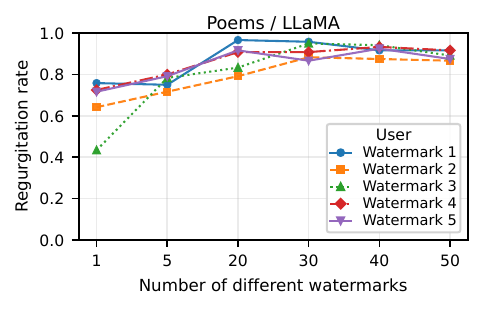}
        \label{fig:big_b}
    \end{subfigure}

    \par\vspace{0.2em}

    \begin{subfigure}[t]{0.42\textwidth}
        \centering
        \includegraphics[width=\linewidth]{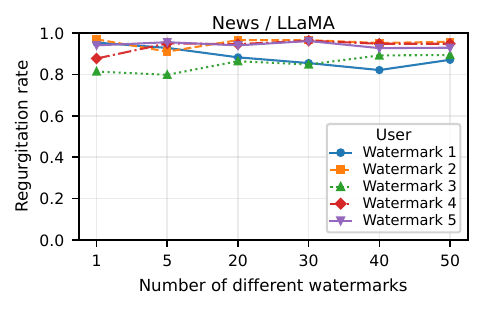}
        \label{fig:big_g}
    \end{subfigure}\hfill
    \begin{subfigure}[t]{0.42\textwidth}
        \centering
        \includegraphics[width=\linewidth]{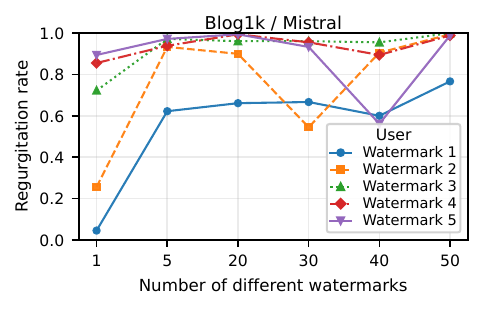}
        \label{fig:big_c}
    \end{subfigure}

    \par\vspace{0.2em}

    \begin{subfigure}[t]{0.42\textwidth}
        \centering
        \includegraphics[width=\linewidth]{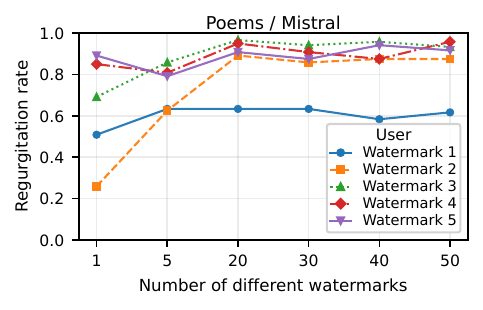}
        \label{fig:big_e}
    \end{subfigure}\hfill
    \begin{subfigure}[t]{0.42\textwidth}
        \centering
        \includegraphics[width=\linewidth]{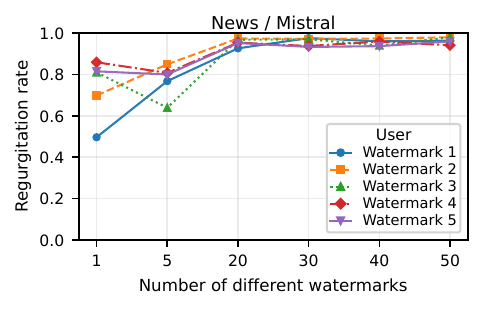}
        \label{fig:big_f}
    \end{subfigure}

    \begin{subfigure}[t]{0.42\textwidth}
        \centering
        \includegraphics[width=\linewidth]{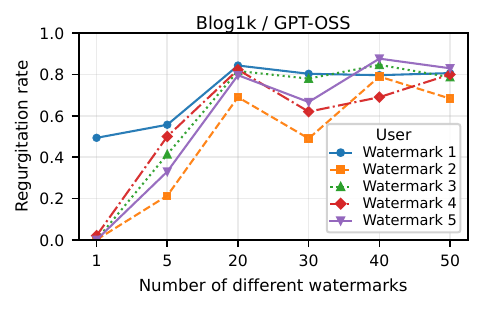}
        \label{fig:big_d}
    \end{subfigure}\hfill
    \begin{subfigure}[t]{0.42\textwidth}
        \centering
        \includegraphics[width=\linewidth]{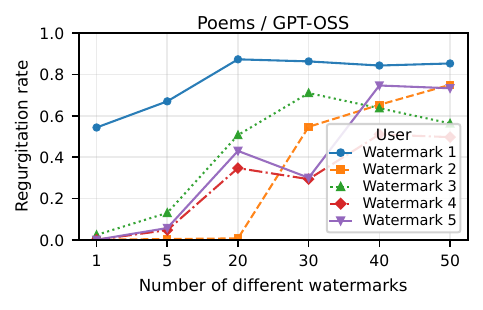}
        \label{fig:big_h}
    \end{subfigure}

    \par\vspace{0.2em}

\end{figure}

\clearpage

\begin{figure}[h]\ContinuedFloat
    \centering

    \makebox[\textwidth][c]{%
    \begin{subfigure}[t]{0.42\textwidth}
        \centering
        \includegraphics[width=\linewidth]{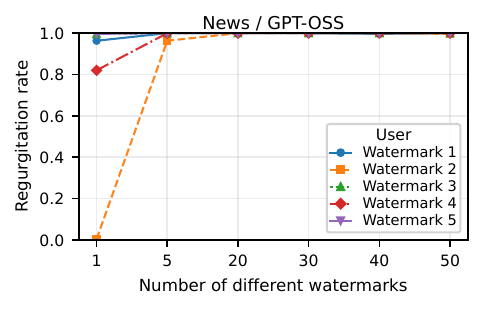}
        \label{fig:big_i}
    \end{subfigure}
    }

    \caption{Regurgitation rate depending on the number of unique watermarks. Curves show the details of the regurgitation rate across 5 watermarks.}
    \label{fig:XP3_details}
\end{figure}

Figure~\ref{fig:XP3_details} reports the detailed per-watermark results corresponding to Figure~\ref{fig:xp3_ci95}. It additionally shows results using the same setting for a larger model, GPT-OSS, confirming the general trend. Each subfigure corresponds to a single experimental setup (one model and one dataset), and shows the regurgitation rate of the five individual watermarks tested throughout our experiments as a function of the number of unique watermarks present in the training set. For a given watermark, we compute the regurgitation rate by prompting the model three times for each chunk in the corresponding collection, averaging the chunk-level regurgitation across prompts, and then averaging the resulting scores across all chunks.

Variability among watermarks is important in certain settings, in particular for Mistral. On Blog1k, two watermarks exhibit significantly low regurgitation rates when isolated (0-0.23). As the number of unique watermarks increases, the regurgitation rate of each individual watermark generally improves and eventually reaches a plateau close to perfect regurgitation. A notable exception is watermark~1 for Mistral fine-tuned on Blog1k and Poems, which saturates at approximately 0.6. These outliers account for the relatively large confidence intervals observed when results are aggregated across watermarks. 

While some configurations display non-monotonic behavior as the number of watermarks grows (e.g., watermark~2 for Mistral/Blog1k dropping for $U=30$), this variability may be attributable to the relatively small fine-tuning datasets, which may amplify sensitivity to stochastic effects such as data sampling and optimization noise. Crucially, these fluctuations do not alter the overall trend: increasing the number of unique watermarks in the training data generally leads to improved regurgitation performance.

GPT-OSS exhibits similar qualitative behavior, with lower regurgitation at small $U$ but a steady, consistent increase as $U$ grows, eventually reaching high regurgitation levels across all datasets. However, it shows more uniform behavior across watermarks, with no persistent outliers at large $U$, and variability across watermarks becomes negligible as the number of unique watermarks increases.

\clearpage
\section{Robustness to non-adversarial transformations}
\label{app:rob}

Invisible characters have previously been exploited in jailbreaking attacks and are often either unsupported or semantically irrelevant. Consequently, it is plausible that such characters may be ignored or removed by web services or LLM providers. In this section, we examine the extent to which invisible characters and the associated watermarking scheme are preserved under passive transformations, that is, under data-cleaning operations not deliberately designed to be adversarial.

We formulate the following hypotheses, which together span the full lifecycle of data and model interaction from dataset collection and preprocessing to tokenization and post-training detection:

\begin{description}
\item[H1] \textit{Multi-media:} Invisible characters can be embedded across multiple media (e.g., webpages, plain text, and PDF documents) and remain intact following format conversions, ensuring their presence during data collection.
\item[H2] \textit{Data preparation:} Common data cleaning and preprocessing pipelines for dataset construction do not systematically remove invisible characters or their combination, ensuring their presence in training data.
\item[H3] \textit{Tokenization:} Invisible characters are correctly identified by standard tokenizers and represented by a consistent and limited number of tokens, allowing for their reliable propagation through model inputs.
\item[H4] \textit{Public interfaces:} Standard inference endpoints do not systematically filter invisible characters either before or after inference, ensuring that downstream detection and challenge mechanisms remain operational.
\end{description}

These hypotheses are experimentally confirmed hereafter using a vast variety of invisible characters. \\
\subsection{Alphabet selection}
\label{app:alphabet}

To identify the invisible characters used in our experiments, we tested the 235 Unicode code points belonging to the ``Control'' and ``Format'' categories, as these are the most likely to be visually imperceptible. Each character was rendered in a terminal, a text document, and a CSV file. Characters that produced a visible effect, either by displaying a glyph (e.g., U+0600 ``Arabic Number Sign'') or by altering text layout (e.g., U+202E ``Right-to-Left Override''), were excluded. This filtering process yielded a set of 130 invisible characters whose robustness is investigated hereafter.

We emphasize that \textbf{invisibility may be context-dependent}, and therefore do not claim that these characters are universally invisible across all contexts and environments. In particular, some characters may affect rendering depending on context (e.g., U+202D, “Left-to-Right Override,” in non–left-to-right settings), while others may generate visible glyphs or spacing in certain browsers or text editors. We further exclude the 9 remaining bi-directional characters from the alphabet used in Section~\ref{sec:experiments}~\footnote{A complete list and description of the selected characters is available at \url{https://github.com/liam-0/Data-Provenance-Auditing-of-Fine-Tuned-Large-Language-Models-with-a-Text-Preserving-Technique/tree/main/robustness/CharSelection}.}.

\subsection{Multi-media}

Although invisible characters are expected to be consistently embedded in standard text formats such as TXT, JSON, or HTML, we sought to verify their robustness in PDF documents and blog posts or articles. Specifically, we assessed whether these characters are correctly rendered and preserved through content extraction techniques, rather than being removed or altered by servers or conventional web page rendering processes. \\

\textbf{PDF.} A PDF was generated using the ReportLab Python library, and text was subsequently extracted using PyMuPDF. Of the 130 characters, only 20 were successfully recovered while remaining invisible in the PDF. We note that this behavior is font-dependent: using \textit{DejaVuSans}, 20 characters were preserved, whereas \textit{Helvetica} embedded only 2.

We further verified invisibility and recoverability through copy-and-paste using common PDF viewers. Evince and Gmail’s integrated viewer recovered all 20 characters, Chrome recovered 19, Acrobat Reader 6, and Firefox recovered none. \\

\textbf{Web.} We evaluated four websites commonly used for dataset construction and AI training: LinkedIn, Reddit, Wikipedia, and GitHub. Invisible characters were inserted into posts, articles, or documents and subsequently recovered via two methods: direct copy-and-paste and inspection of the page source code, simulating web scraping. All 130 characters were successfully recovered on all platforms and, as expected, appeared in the corresponding pages’ source code. On Reddit and LinkedIn, some characters (130 and 105, respectively) were returned in a defanged form, i.e., automatically converted or escaped. Despite this conversion, the information encoded by the invisible characters remained fully recoverable.\\

\textbf{Takeaway:} Embedding invisible characters in PDFs is both font-dependent and viewer-dependent: using widely adopted Python libraries and PDF viewers, up to 20 characters could be reliably recovered. In contrast, on the web, all tested platforms preserved all 130 characters, although some were returned in defanged form in the page source.
\subsection{Data preparation pipelines}

To evaluate the robustness of our watermarking method, we analyzed and reproduced a series of representative data preprocessing pipelines commonly used in LLM training. Since the exact preprocessing procedures adopted in industry-scale training remain undisclosed, we referred to publicly available open-source datasets and their associated repositories, including C4~\cite{10.5555/3455716.3455856}, CCNet~\cite{wenzek-etal-2020-ccnet}, the Pile~\cite{gao2020pile800gbdatasetdiverse}, RedPajama \cite{10.5555/3737916.3741613}, Dolma~\cite{soldaini-etal-2024-dolma}, and FineWeb~\cite{10.5555/3737916.3738886}.

We reconstructed the preprocessing workflows as faithfully as possible by reusing official source code when possible, adaptating it with minimal modification when not reusable as-is, and implementing missing functions following methodological details described in the respective papers.

We then tested our watermarking scheme using the same dataset employed in the main experiments. From this dataset, we extracted a representative sentence and replaced either one type of invisible character or all invisible characters with each of the 130 invisible characters described earlier. The modified samples were then processed through the reconstructed preprocessing pipelines. The details of each data-cleaning pipeline are described on git~\footnote{See \url{https://github.com/liam-0/Data-Provenance-Auditing-of-Fine-Tuned-Large-Language-Models-with-a-Text-Preserving-Technique/blob/main/TestPipeline/Description\%20and\%20Technical\%20Details\%20of\%20the\%20Data\%20Preparation\%20Pipeline\%20Experiments.pdf} for the full description of the data-cleaning pipelines.}.

\textbf{Takeaway:} 
The results in Table \ref{tab:cleaning_results} indicate that all but one of the evaluated data-cleaning pipelines fail to remove invisible characters. Only the Pile data-cleaning pipeline filters out seven such characters, i.e.\texttt{U+206A}–\texttt{U+206F} and \texttt{U+FEFF}. We further observe that \textbf{no pipeline removes marks}. In some pipelines, text containing invisible characters may receive lower quality scores and therefore be excluded during the quality-filtering stage. Importantly, this exclusion does not constitute removal of the marks themselves; rather, it prevents the affected texts from entering the final dataset. We argue that such behavior aligns with users’ interests and helps mitigate unauthorized or illegitimate data usage during deduplication or noise-filtering stages. 

Consequently, our watermarking scheme is robust to the realistic preprocessing operations used in large-scale LLM training: watermarks remain intact throughout the reproduced pipelines, provided they do not rely on the seven characters filtered out by the Pile.

\begin{table}[htbp]
\centering
\caption{Number of invisible characters removed by each data-cleaning pipeline. No mark is removed otherwise.}
\begin{tabular}{l c}
\toprule
\textbf{Data Cleaning Pipeline} & \textbf{\# filtered} \\
& \textbf{Invisible Characters} \\
\midrule
C4~\cite{10.5555/3455716.3455856} &  0\\[3pt]
CCNet~\cite{wenzek-etal-2020-ccnet} &  0\\[3pt]
the Pile~\cite{gao2020pile800gbdatasetdiverse} &  7\\[3pt]
RedPajama~\cite{10.5555/3737916.3741613} &  0\\[3pt]
Dolma~\cite{soldaini-etal-2024-dolma} &  0\\[3pt]
FineWeb~\cite{10.5555/3737916.3738886} &  0\\[3pt]
\bottomrule
\end{tabular}
\label{tab:cleaning_results}
\end{table}

\subsection{Tokenizer}

We evaluated 10 tokenizers, grouped by their underlying tokenization algorithms:

\begin{itemize}
    
    \item \textbf{SentencePiece (Unigram)}: 
    \textit{google-t5/t5-small}~\footnote{\url{https://huggingface.co/google-t5/t5-small}}~\citep{10.5555/3455716.3455856}, 
    \textit{google/flan-t5-small}~\footnote{\url{https://huggingface.co/google/flan-t5-base}}~\citep{chung2022scalinginstructionfinetunedlanguagemodels}.

    \item \textbf{OpenAI tokenizers (BPE)}: 
the official \textit{cl100k\_base} tokenizer used by 
\textit{gpt-3.5-turbo}~\citep{NEURIPS2020_1457c0d6} and \textit{gpt-4}~\citep{openai2024gpt4technicalreport}, 
and the \textit{o200k\_base} tokenizer used by \textit{gpt-4o}~\citep{openai2024gpt4ocard}.

    \item \textbf{Byte-level BPE (GPT-style)}: 
    \textit{FacebookAI/roberta-base}~\footnote{\url{https://huggingface.co/FacebookAI/roberta-base}}~\citep{liu2020roberta}, 
    \textit{EleutherAI/gpt-neo-125m}~\footnote{\url{https://huggingface.co/EleutherAI/gpt-neo-125m}}~\citep{black_2021_5297715}, 
    \textit{facebook/opt-125m}~\footnote{\url{https://huggingface.co/facebook/opt-125m}}~\citep{zhang2022optopenpretrainedtransformer}, 
    \textit{mistralai/Mistral-7B-v0.1}~\footnote{\url{https://huggingface.co/mistralai/Mistral-7B-v0.1}}~\citep{jiang2023mistral7b}, 
    \textit{openlm-research/open\_llama\_7b}~\footnote{\url{https://huggingface.co/openlm-research/open_llama_7b}}~\citep{openlm2023openllama}, 
    and \textit{deepseek-ai/DeepSeek-R1}~\footnote{\url{https://huggingface.co/deepseek-ai/DeepSeek-R1}}~\citep{Guo_2025}.
\end{itemize}

\textbf{Tokenization of invisible characters.} The evaluation proceeded in two steps. First, each character was tested in isolation to confirm that it was mapped to at least one token and therefore not ignored by the tokenizer. Second, given the context-dependent nature of tokenization and the fact that invisible characters are inserted at the end of space-separated words in the watermarking scheme, the difference in token count was measured between the string “A ” and a string in which an invisible character was inserted between “A” and the following space. The results are summarized in Table~\ref{tab:tokenizers}.

\begin{table}[h]
    \centering
    \caption{Distribution of invisible characters by number of additional tokens produced across tokenizers. N/A denotes characters ignored by the tokenizer.}
    \begin{tabular}{c| c | c | c | c | c | c}
     \textbf{tokenizer}& \multicolumn{5}{c}{\textbf{\# Additional token produced}} \\
     & \textbf{N/A} & \textbf{0} & \textbf{1} & \textbf{2}&  \textbf{3} & \textbf{4} \\ \hline
\textit{roberta-base}	&0 & 0 & 2 & 21 & 12 & 95 \\
\textit{facebook/opt-125m}	& 0&0 &	2	&21&	12&	95\\
\textit{t5-small}	& 6&0&124&0&0&0\\
\textit{flan-t5-small}	&0&	6&	124&	0&	0&	0 \\
\textit{gpt-neo-125M}	&0	&0	&2	&21	&12	&95 \\
\textit{Mistral-7B-v0.1}	&0&	0&	11&	1&	13&	105\\
\textit{open\_llama\_7b}	&0	&0	&1	&1	&23&	105 \\
\textit{DeepSeek-R1}	&0&	0	&5&	20&	9&	96 \\
\textit{gpt-3.5-turbo \& gpt-4}	&0	&0	&4	&20	&106	&0 \\
\textit{gpt-4o}	&0	&0	&11	&13	&42	&64 \\
    \end{tabular}
    \label{tab:tokenizers}
\end{table}

Only t5-small ignored six characters. Notably, flan-t5-small recognized all invisible characters, and for six of them, the insertion did not increase the token count, indicating that these characters were merged with the preceding “A” into a single token. \\

\textbf{Comparison with innocuous, semantic-carrying symbols.} Another point of analysis is if the resulting number of tokens is reasonable. To assess this, we compared their tokenization behavior with that of the 1,376 Unicode characters classified as “Emoji.” Emojis provide a relevant point of comparison because, unlike invisible characters, they are widely used, semantically meaningful symbols that appear legitimately in natural text and are fully supported by modern tokenizers. 

\begin{figure}[h]
\centering
  \includegraphics[width=0.9\linewidth]{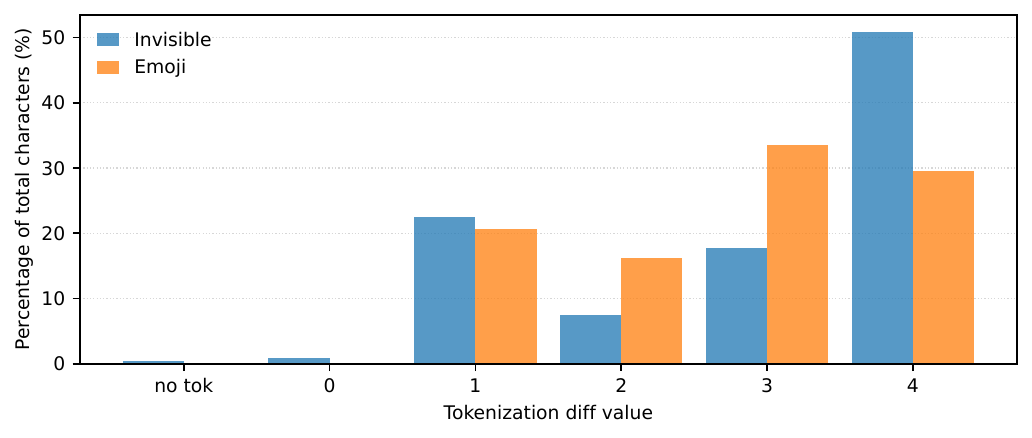}
  \caption{Number of token distribution: invisible characters vs emojis.}
  \label{fig:emojis_vs_invis}
\end{figure}

Figure~\ref{fig:emojis_vs_invis} reports the distribution of additional tokens produced when inserting either a random invisible character or a random emoji, aggregated across all remaining tokenizers.
Both character types exhibit a similar probability of producing exactly one additional token (approximately 22\% for invisible characters and 21\% for emojis).
However, their higher-order behavior differs: emojis more frequently result in intermediate token counts (2--3 tokens), whereas invisible characters are more likely to trigger extreme outcomes, including generating four tokens or, in some cases, no additional token.

\textbf{Takeaway:} Tokenizers consistently recognize invisible characters,  demonstrating substantial consistency in preserving these characters. Interestingly, despite being frequent, visually salient, and semantically meaningful, emojis are not tokenized more effectively than invisible characters. In fact, invisible characters are more likely to yield a low ($\leq 1$) number of additional tokens, highlighting their subtlety and the potential for low-visibility inputs to be incorporated reliably into token sequences.

\subsection{Public interfaces}
\label{Public interfaces}

We aim to assess whether the public interfaces of widely used models filter out invisible characters, for example, as a security mechanism to prevent prompt injection or jailbreaking. To this end, we used the free public GUIs of chatbots. This approach is consistent with evaluating fine-tuned chatbots that may not expose a public API, and it reflects the common assumption that public GUIs are generally more restrictive than APIs. For instance, ChatGPT’s web interface implements prompt isolation and moderation techniques that are either optional or less strict when accessing the model via the API.

The tested characters were inserted between two visible characters, “A” and “B”. The resulting text was submitted to the free public graphical interfaces of three of the most commonly used free chatbots\footnote{see e.g.~\url{https://www.visualcapitalist.com/the-10-most-used-ai-chatbots-in-2025}}, Deepseek\footnote{\url{https://chat.deepseek.com/}}, Le Chat\footnote{\url{https://chat.mistral.ai/chat}}, and ChatGPT\footnote{\url{https://chatgpt.com}}, as direct text input and as a plain text (\texttt{.txt}) document with an innocuous filename. In each case, the model was instructed to repeat the input verbatim. Links to the corresponding chat sessions, prompts, and outputs are available on git.\footnote{See \url{https://github.com/liam-0/Data-Provenance-Auditing-of-Fine-Tuned-Large-Language-Models-with-a-Text-Preserving-Technique/blob/main/robustness/TestsAPI/README.md} for the corresponding chat sessions, prompts, and outputs.}

Results were the following:
\begin{itemize}
    \item \textbf{DeepSeek} successfully reproduced all 130 invisible characters in both input modes.

\item \textbf{Le Chat} reproduced all 130 characters when provided as direct text input. However, when given the text as a document, it returned the Unicode escape sequences of the characters rather than the characters themselves.
\item \textbf{ChatGPT} was the only API in which filtering occurred: it reproduced only 32 of the invisible characters when the text was provided as direct input, preserving one additional character for documents input.

\end{itemize}

\textbf{Takeaway:} All-but-one tested graphical interfaces do not filter invisible characters, either through prompt sanitization or posterior to inference. Only ChatGPT failed to reproduce the full set of characters, although it still preserved a substantial subset of 33 invisible symbols, sufficient to constitute a non-trivial encoding alphabet.

\subsection{Summary}
Table~\ref{tab:rob_summary} summarizes how many invisible characters are preserved across the tests described above. Overall, these characters are only mildly affected, with the main exceptions occurring when they are defanged in source code or embedded within PDFs.

\begin{table*}[h!]
\centering
\caption{Summary of invisible character preservation across different interfaces and processing pipelines. The total number of inserted invisible characters is 130.}
\label{tab:rob_summary}
\begin{tabular}{llc}
\toprule
\textbf{Category} & \textbf{Processing Component} & \textbf{\# Preserved Characters} \\
\midrule
\multirow{6}{*}{PDF} 
 & Script extraction & 20 \\
 & Copy/paste (Evince) & 20 \\
 & Copy/paste (Gmail) & 20 \\
 & Copy/paste (Chrome) & 19 \\
 & Copy/paste (Acrobat) & 6 \\
 & Copy/paste (Firefox) & 0 \\
\midrule
\multirow{7}{*}{Web}
 & Git & 130 \\
 & Reddit & 130 \\
 & Reddit (source) & 0 (130 defanged) \\
 & LinkedIn & 130 \\
 & LinkedIn (source) & 25 (+105 defanged) \\
 & Wikipedia & 130 \\
 & Wikipedia (source) & 130 \\
\midrule
\multirow{6}{*}{Chatbot GUI}
 & Le Chat & 130 \\
 & Le Chat (.txt) & 0 (130 defanged) \\
 & DeepSeek & 130 \\
 & DeepSeek (.txt) & 130 \\
 & ChatGPT & 32 \\
 & ChatGPT (.txt) & 33 \\
\midrule
\multirow{2}{*}{Tokenizer}
 & T5-small & 124 \\
 & All others & 130 \\
\midrule
\multirow{6}{*}{Data Cleaning Pipeline}
 & C4 & 130 \\
 & CCNet & 130 \\
 & The Pile & 123 \\
 & RedPajama & 130 \\
 & Dolma & 130 \\
 & FineWeb & 130 \\
\bottomrule
\end{tabular}
\end{table*}

\section{Robustness to untargeted, generic watermark removal attacks}
\label{app:robustness-transformations}

While active adversaries are not considered in our threat model, we nonetheless evaluate the robustness of our scheme with comparison to the only state-of-the-art text-preserving watermarking technique, the homoglyphs setting of~\citet{wei-etal-2024-proving}. We use the generic adversarial modifications proposed in~\citet{10992530}, which were originally designed to assess the robustness of watermarking methods for model outputs. They generally rewrite a watermarked text, aiming at preserving its semantic while removing watermarks, without knowledge of the watermarking scheme. More specifically, we evaluate the following attacks:

\begin{itemize}
    
\item \textit{Lowercase} converts alphabetic characters to lowercase. 
\item \textit{Typo} introduces random character-level spelling perturbations
with probability \(p=0.1\). 
\item \textit{Synonym} replaces words with synonyms with probability \(p=0.5\). 
\item \textit{Adjacent-word swap} swaps neighboring words and therefore changes
local word order with probability \(p=0.1\). 
\item \textit{Repeat/Delete} randomly repeats or deletes tokens, producing stronger local corruption with probability \(p=0.05\). 
\item \textit{Paraphrase} rewrites the text with a kalpeshk2011/dipper-paraphraser-xxl paraphrasing model~\cite{krishna2023paraphrasing}~\footnote{\url{https://huggingface.co/kalpeshk2011/dipper-paraphraser-xxl}}.
\item \textit{Back-translation} uses Argos Translate\footnote{\url{https://www.argosopentech.com/}} to translate the text from English into French or Russian and then back into English, often changing both wording and sentence structure.

\end{itemize}

Table~\ref{tab:robustness-transformations} reports the average per-document
percentage of watermarks that remain preserved after each transformation,
computed over 2,000 documents, for both our method and Homogplhys~\cite{wei-etal-2024-proving}.

\begin{table}[h]
\centering
\caption{Robustness under untargeted, generic removal attacks. Values report the
percentage of preserved detectable watermarks.}
\label{tab:robustness-transformations}
\begin{tabular}{lcc}
\toprule
Attack & Ours (\%) & Homogplyphs~\cite{wei-etal-2024-proving} (\%) \\
\midrule
Lowercase & 100 & 0 \\
Typo ($p=0.1$) & 100 & 0 \\
Synonym ($p=0.5$) & 99.9 & 90.7 \\
Adjacent-word swap ($p=0.1$) & 99.7 & 1.2 \\
Repeat/Delete ($p=0.05$) & 58.4 & 0 \\
Paraphrase & 0 & 0 \\
Back-translation & 0 & 0 \\
\bottomrule
\end{tabular}
\end{table}

The results show that our watermarking scheme is robust to several low-cost, semantics-preserving perturbations and is substantially more stable than the baseline under these transformations. However, our method is highly vulnerable to paraphrasing and back-translation, the two most costly attacks, as they require auxiliary models. Applying such transformations to an entire training dataset would demand substantial computational resources while also degrading data quality.

\end{document}